\documentclass[]{interact}

\usepackage{epstopdf}
\usepackage{subfigure}
\usepackage{upgreek}
\usepackage{ifthen}
\usepackage[innercaption,raggedright,rightcaption]{sidecap}

\newcommand{\comments}[1]{}

\usepackage[numbers,sort&compress,merge]{natbib}
\bibpunct[, ]{(}{)}{,}{n}{,}{,}

\theoremstyle{plain}

\theoremstyle{definition}

\theoremstyle{remark}

\renewcommand{\text}[1]{\textrm{\tiny{#1}}}

\begin{document}
                  
\articletype{JMO MANUSCRIPT - SINUCO-LEON and FROMHOLD }

\title{Atom chips with free-standing two-dimensional electron gases: advantages and challenges}

\author{
\name{G. A. Sinuco-Le\'on\textsuperscript{a}\thanks{CONTACT G.A. Sinuco-Leon. Email: g.sinuco@sussex.ac.uk}, Peter Kr\"uger \textsuperscript{a} and T.M. Fromhold \textsuperscript{b}}
\affil{\textsuperscript{a}Department of Physics and Astronomy, University of Sussex, Brighton, BN1 9QH, UK. \\ \textsuperscript{b}School of Physics and
  Astronomy, University of Nottingham, Nottingham, NG7 2RD, UK.}
}

\maketitle

\begin{abstract}
In this work we consider the advantages and challenges of using free-standing two-dimensional electron gases (2DEG) as active components in atom chips for manipulating  ultracold ensembles of alkali atoms. We calculate  trapping parameters achievable with typical high-mobility 2DEGs in an atom chip configuration, and identify advantages of this system for trapping atoms at sub-micron distances from the atom chip. We show how the sensitivity of atomic gases to magnetic field inhomogeneity can be exploited for controlling the atoms with quantum electronic devices and, conversely,  using the atoms to probe the structural and transport properties of semiconductor devices.  
\end{abstract}


\section{Introduction}
\label{sec:introduction}
Atom chip technology has become a mature engineering tool for trapping, manipulating and controlling ultracold atomic matter \cite{FifteenYearsFolman}. Developed initially to address neutral alkali atoms \cite{reichel1999atomic}, atom chips now find applications in a number of platforms for quantum physics such as trapped-ions \cite{DeMotte2016}, trapped-electrons \cite{cridland2016single} and ultracold molecules \cite{PhysRevLett.100.153003}. They are also emerging as promising tools for manipulating antimatter \cite{FifteenYearsFolman}. Advances in material science and microfabrication techniques have allowed the integration of a number of devices and materials into the atom chip, opening the prospect of creating hybrid quantum systems that exploit the complementary capabilities of atomic matter-waves and solid-state devices for applications in quantum information processing, sensing and metrology.  

Such applications require short distances between the trapped and trapping elements. However, atom-surface proximity effects have restricted  most atom chip experiments to date to distances exceeding $1~\mu$m, with the exception of sub-micron trapping achieved using evanescent light fields \cite{PhysRevLett.90.173001, PhysRevLett.104.203603}. The common use of metallic conductors in atom chips have limited the  miniaturisation of the potential landscape, since atomic ensembles become disturbed by intense Johnson-Nyquist noise \cite{noise1,noise2,noise3,zhang2005relevance}, strong atom-surface Casimir-Polder (CP) attraction \cite{noise3} and defect-induced fluctuations in trapping potentials  \cite{PRL92_076802,PhysRevA.76.063621,rough1,rough2,rough3}. A number of proposals have been put forward to overcome these issues \cite{Folman_sub,doi:10.1063/1.2219397,PhysRevA.76.063621,PRL98_263201}, drawing tools from nanofabrication and low-frequency dressing of atomic states. However, there is currently no single approach that comprehensively overcomes all of these difficulties.

The sensitivity of ultracold atoms to magnetic field fluctuations has, conversely, been exploited to develop atomic Bose-Einstein condensate (BEC) microscopy and investigate the properties of the atom chip components \cite{APL88_264103,Nature_wildermuth,PhysRevApplied.7.034026}. The ground breaking feature of BEC microscopy lies in its ability to sense static and AC magnetic fields, combining high-sensitivity with high-spatial resolution and single-shot imaging of large areas \cite{JPCS19_56,JAP65_361,APL88_264103,PhysRevApplied.7.034026,Nature_wildermuth,aigner2008long}. Further developing this technique will provide imaging access to a number of physical phenomena observed in solid-state devices (e.g. quantisation of conductance, weak and strong electronic localisation), about which our current knowledge usually derives from indirect transport measurements \cite{FifteenYearsFolman, Bastard}. 

\begin{figure}
  \centering   
  \includegraphics[width=14cm]{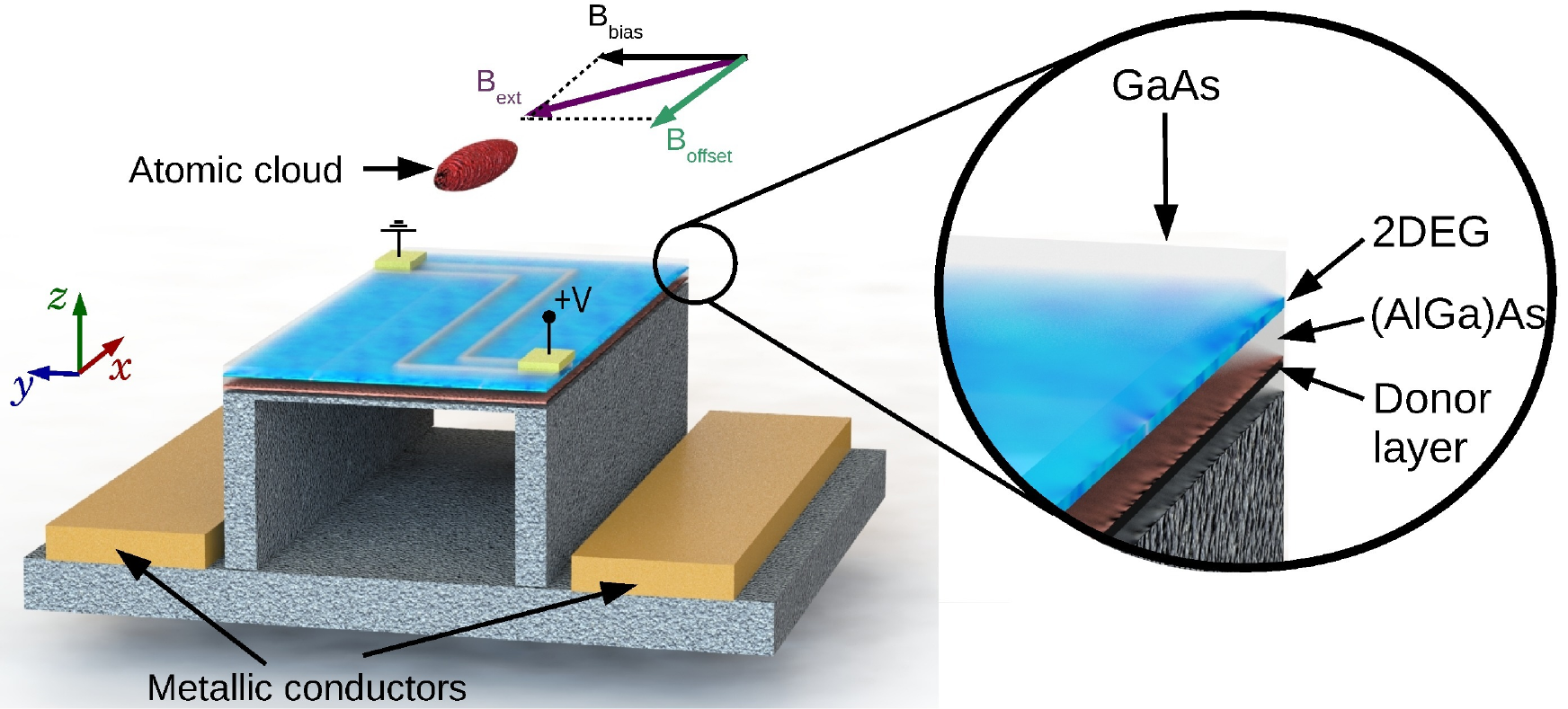}
  \caption{\label{geometry} Schematic diagram of a hybrid 2DEG-atom chip, showing a BEC (red) trapped in the vicinity of a free-standing heterostructure containing  a 2DEG (upper layered structure). The magnetic trapping potential can be produced either by current-carrying metallic conductors (two of them shown as light brown slabs), by current-carrying channels defined in the 2DEG (Z-shaped enclosed area in the 2DEG, bounded by the grey Z-shaped boundaries) or by a combination of the two. Metallic gates deposited on top of the heterostructure's uppermost layer (small light-yellow squares) control the current in the 2DEG channel. Magenta arrow represents an applied magnetic field, $B_{\text{ext}}$, with components $B_{\text{bias}}$ (black arrow) and $B_{\text{offset}}$ (green arrow), used to control the trap position and tightness. Inset: Close up showing the layers of the heterostructure.} 
\end{figure}

Here we consider atom chips that include conducting channels defined in high mobility two-dimensional electron gases (2DEGs) in free-standing heterostructures \cite{Bastard,graphene,judd2011quantum}.  A schematic diagram of such a hybrid atom chip is shown in figure \ref{geometry}, where the central feature is a free-standing heterostructure (blue), whose layered structure is shown enlarged on the right. To demonstrate the advantages of using 2DEGs in atom chips, we focus on two possible functions of this device.  Firstly, we study how the magnetic field generated by electric currents in conductors patterned in a 2DEG can be used to trap and manipulate atomic BECs. Secondly, we examine how BEC microscopy can provide us with information about transport processes in 2DEGs.

We consider a typical free-standing $\delta$-doped GaAs/(AlGa)As heterostructure containing a high-mobility 2DEG, shown in figure \ref{geometry}. For all calculations in this work, we consider a 2DEG with a mean electron density of $n = 3.3 \times 10^{15}~$m$^{-2}$ and electron mobility of $\mu = 140$ m$^2$V$^{-1}$s$^{-1}$ at liquid helium temperature \cite{PRB47_2233}, which corresponds to a conductivity $\sigma = e n \mu = 0.074~\Omega^{-1}$. Also, the plane of the 2DEG is separated from a layer of ionised donors by $d=52.9~$nm. Details of the model used to calculate currents in the 2DEG are given in Appendices \ref{sec:2DEG} and \ref{sec:field}. We evaluate the performance of the 2DEG atom chip by considering magnetic trapping of $^{87}$Rb in the hyperfine state $\left\vert F=2,m_F=2 \right\rangle$.

Our numerical investigations suggest that this type of atom chip has a number of advantages derived from the flexibility of tailoring the transport properties of the 2DEG, and, in particular, offers a favourable platform for submicron trapping of atomic matter-waves and coupling to quantum electronic devices. 

\section{Advantages of using 2DEGs in atom chips: reduced spatial noise and long life-times}
\label{sec:overview}
Two-dimensional electron gases in semiconductor heterostructures are essential components of a major class of modern quantum electronic devices \cite{pepper}. In addition to being a superb test bed for investigating fundamental physical phenomena, 2DEGs currently have important technological and industrial applications, including setting the international standard of resistance and enabling high-mobility transistors for mobile communication devices. Despite their transformative role in electronics, there has been little discussion of creating hybrid quantum systems based on current-carrying 2DEGs and near-surface trapped ultracold atomic matter \cite{OurNJP, PhysRevA.83.021401,PhysRevB.92.245439}. Here, we discuss the challenges and potential advantages that such hybrid systems offer for manipulating atomic matter-waves and in materials research.

Strong coupling between neutral ultra-cold atomic matter and quantum electronic devices in an atom chip architecture requires an atom-surface separation below $1~\mu$m, such that the atoms and material charge carriers couple via their magnetic moments or dynamical electric dipoles  \cite{0953-4075-46-24-245502}. However, achieving such small separation requires a number of challenges to be overcome. Firstly, at sub-micron distances an intense atom-surface attractive Casimir-Polder force dominates, making it difficult to create magnetic trapping potentials that prevent the atom cloud from collapsing onto the chip surface  \cite{noise3,judd2011quantum}.  Secondly, as the atoms get closer to a surface, their coupling to the electromagnetic Johnson-Nyquist noise produced by thermal motion of conduction electrons becomes strong,  leading to a reduction in the lifetime of trappable atomic states \cite{noise1,noise3,Henkel1999}. Finally, effects originating from fabrication defects of the atom chip components magnify as the atom-surface separation is reduced \cite{PhysRevA.76.063621,rough3}, making it difficult to define and control smooth atomic potential landscapes.

All of these challenges can be overcome by using free-standing atom chips containing two-dimensional electron gases, such as those present in semiconducting heterostructures, doped SiN ultrathin layers, or graphene sheets \cite{graphene,judd2011quantum}. These systems are favourable since their transport properties can be statically or dynamically tuned by a number of experimental tools, including tailored fabrication, active control of the operating temperature, and partial illumination. Also, free-standing ultra-thin membranes will exert only weak attractive forces on neighbouring atoms, due to the significantly reduced volume of dielectric material acting on the atoms \cite{Henkel1999,PhysRevA.80.032901,sernelius2015casimir,judd2011quantum}. Finally,  the intrinsic low electron density of 2DEGs greatly reduces the Johnson-Nyquist noise in their immediate vicinity, from which it follows that the lifetime of trappable atomic states at submicron distances from typical 2DEGs can become of the order of a few hundreds of seconds (see section \ref{sec:life_time}) \cite{PhysRevA.83.021401,zhang2005relevance}. 

Another advantage of using 2DEGs over the metallic conductors usually employed in atom chips is the negligible magnetic field fluctuations resulting from surface and edge irregularities,  which are reduced due to the extremely thin nature of the 2DEG ($\sim 15~$nm thick) and high accuracy of fabrication methods available for defining conducting channels in 2DEGs (e.g. the typical resolution of ion implanting is $\sim 10~$nm \cite{ion_beam2,ion_beam3}). In addition, as shown below in section \ref{sec:fieldinhomegneities}, irregular electronic flow in a 2DEG can be controlled by imprinting a periodic pattern in it via optical illumination \cite{xray}, etching, or ion-implanting the 2DEG \cite{ion_beam3}. Such patterning results in an exponential decay of the field inhomogeneity when moving away from the 2DEG  \cite{PhysRevA.83.021401}, with a decay length equal to the period of the pattern. Thus, by imprinting a submicron periodic pattern, the root-mean-square fluctuations of the magnetic field can be up to $3$ orders of magnitude weaker near a 2DEG than in the vicinity of metallic wires, at distances of  $\sim 1~\mu$m \cite{PRL92_076802,EPJD32_171}. 
 
Collectively, the reduced Johnson-Nyquist noise, weak atom-surface attraction, and routes to defining smooth field distributions, make free-standing membranes with 2DEGs ideal for producing smooth near-surface magnetic traps with long lifetimes, as required to create hybrid cold-atom/quantum electronic systems and devices.

\subsection{Life-times of magnetically trappable atomic states near a 2DEG}
\label{sec:life_time}
In many atom chip configurations, the atomic potential energy is defined by a magnetic field landscape produced by microfabricated permanent magnets or current-carrying conductors \cite{RMP79_235}. Typically, this approach produces a confining potential for a subset of Zeeman split states of the ground state manifold and, therefore, transitions between those states reduce the number of trapped atoms. At short separations between the atoms and macroscopic chip elements, transitions between hyperfine states are enhanced by coupling to thermal electromagnetic modes (in the form of Johnson-Nyquist noise) surrounding the atom chip structure.

The scale of this effect is characterised by the spectral density of magnetic field fluctuations near the atom chip surface, which depends on the chip material, geometry and temperature  \cite{PRA56_2451, RMP79_235}.  In particular, transition rates between Zeeman states near a thin layer of conducting material, such as the 2DEG, depend on its conductivity, $\sigma$, and the atom-surface separation, $z$,  according to \cite{PhysRevA.72.042901,noise1}:
\begin{equation}
\Gamma (z) = \frac{9}{64} \frac{n_{\text{th}} +
  1}{\tau_0}  \left(\frac{c}{\omega_{\text{fi}}}\right)^3 \mu_0 \omega_{\text{fi}}\sigma \frac{1}{z^2},
\label{life_time_reduced}
\end{equation}
where $\omega_{\text{fi}}$ is the transition frequency (here determined by the Zeeman splitting), $n_{\text{th}}$ is the thermal occupancy Bose factor of electromagnetic modes with energy  $\hbar \omega_{\text{fi}}$, $\tau_0$ is the lifetime of the atomic state in free space, $c$ is the speed of light and $\mu_0$ is the magnetic permittivity of vacuum.
\begin{figure}[!htb]
  \centering
  \includegraphics[width=8.0cm]{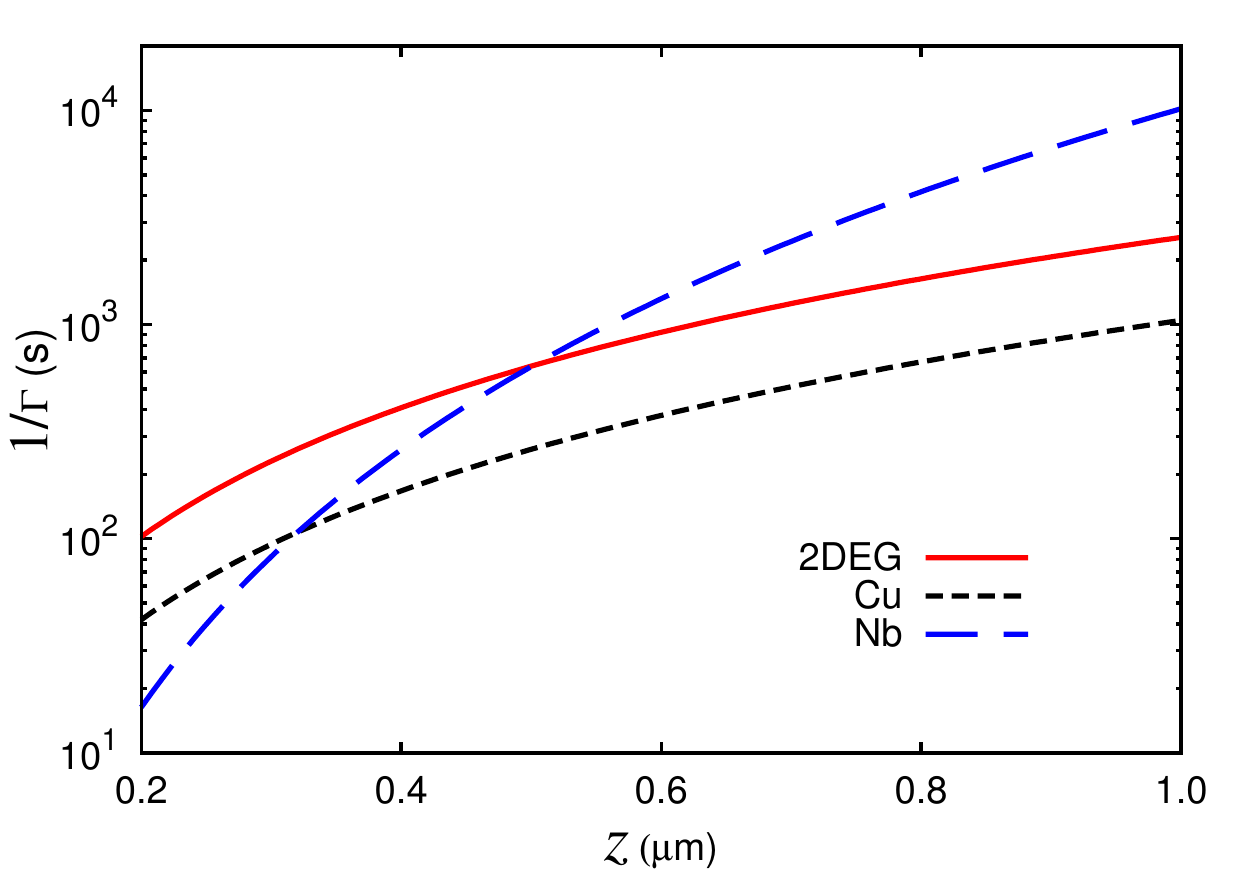}
  \caption{\label{lifetime} Lifetime, $1/\Gamma$, calculated for the hyperfine ground state $\left|F=2,m_F=2\right\rangle$ of $^{87}$Rb above a 2DEG (solid/red curve), a $10~$nm thick layer of copper (short-dashed/black curve) and a thick slab of superconducting niobium (dashed/blue curve).}
\end{figure}

In figure \ref{lifetime} we compare the lifetime, $1/\Gamma$,  of the atomic state $\left|F=2,m_F=2\right\rangle$ of $^{87}$Rb near three materials: a high-mobility 2DEG as specified in section \ref{sec:introduction}, a $10~$nm thick layer of copper at room temperature and a superconducting slab of niobium at $T=4.2~$K.  In all three cases, the Zeeman splitting frequency is set to  $\omega_{\text{fi}}/2 \pi = 1~$MHz. Note that the expected atomic lifetime above a 2DEG is much larger than above a copper layer of similar thickness ($10~$nm). This is because the small conductivity and quasi-two dimensional character of the 2DEG makes its electromagnetic noise spectrum significantly weaker.

For comparison, the lifetime of atomic states above a superconducting slab of niobium can be estimated using  Eq. (9) of \cite{PRL97_070401}:
\begin{equation}
\Gamma_{\text{SC}}(z) =  \frac{1}{\tau_0} (n_{\text{th}}+1) \left[ 1 + 2 \left(\frac{3}{4}\right)^3 \left(\frac{c}{\omega_{\text{fi}}}\right)^3\frac{\lambda_{\text{L}}^3(T)}{ \delta (T)^2} \frac{1}{z^4}\right],
\end{equation}
where $\lambda_{\text{L}}$ is the London penetration length and $\delta(T)$ is the temperature-dependent skin depth of normal conducting electrons. It has been shown in \cite{PRL97_070401} that the efficient screening properties of superconducting materials make the life-time decay with a higher power of the distance compared with normal metallic materials. Figure 2 shows that the lifetime near a 2DEG is about four times shorter than near the superconducting surface at atom-surface separations of $z\approx 1~\mu$m. This trend reverses when these two length scales are comparable and the stronger variation of $\Gamma$ with $z$ for the superconducting case ($1/z^4$ vs. $1/z^2$) dominates the atomic lifetime. Consequently, the lifetime near the 2DEG exceeds that for the superconducting slab when $z \le 0.5~\mu$m,  making the 2DEG more favourable for submicron trapping.

\subsection{Control of inhomogeneity in 2DEGs}
\label{sec:fieldinhomegneities}

Very early in the development of atom chip technology, magnetic field inhomogeneity was identified as responsible for the spatial fragmentation of ultracold atomic ensembles trapped in the vicinity of current-carrying conductors \cite{leanhardt2002propagation, rough1,rough3,Nature_wildermuth}. This is because the electric current produces magnetic field fluctuations that originate from the meandering of the trajectories of free-charges. In the metallic conductors used in atom chips, material defects and edge imperfections cause modulation of the magnetic field produced when an electric current flows \cite{rough1,rough3,PRL92_076802,PhysRevA.76.063621}. Governed by the Biot-Savart law, such modulations of the magnetic field increase as the distance to the conductor decreases, becoming the dominant feature of the magnetic field profile at separations of the order of the length scale of the imperfection \cite{rough1,rough3,PRL92_076802}.  

In the case of 2DEGs in semiconductor heterostructures, the main source of defects that affect electronic transport is the electrostatic interaction between the electrons and ionised donors (see inset of figure \ref{geometry}). In $\delta$-doped heterostructures, the ionised donors are distributed in a thin layer (red layer in figure \ref{geometry}) separated from the 2DEG by $d \sim 10$s nm-$100$s nm. The ionised donor density profile has a mean density similar to the charge carrier density in the 2DEG and spatial-fluctuations $n_d(x,y,z=-d)$, where $z=0$ defines the 2DEG plane. This inhomogeneous distribution of ions creates an electrostatic potential energy landscape for the electrons in the 2DEG, $\Phi(x,y)$, which disturbs their trajectories when a uniform electric field is applied in the plane of the 2DEG. The resulting perturbed trajectories of electrons in high-mobility 2DEGs can be calculated using a linear screening approximation to calculate $\Phi(x,y)$, as we explain in Appendix \ref{sec:2DEG}. 

The inhomogeneity of the magnetic field produced when a small current flows in a 2DEG is determined by the density-density correlation of the donor distribution (see Appendix \ref{sec:2DEG}) \cite{Mico,xray}. Semiconducting heterostructures offer the opportunity to manipulate the donor distribution in several ways, including thermal cycling, sample illumination and ion-implanting. Such manipulation enables reduction of the inhomogeneity intrinsic to a random distribution of donors. In particular, periodic modulation of the donor density leads to an exponential suppression of the high-spatial frequency components of the current and the corresponding field inhomogeneity \cite{PhysRevA.83.021401}.  

\begin{figure}[!h]
  \centering 
  \includegraphics[width=9.0cm]{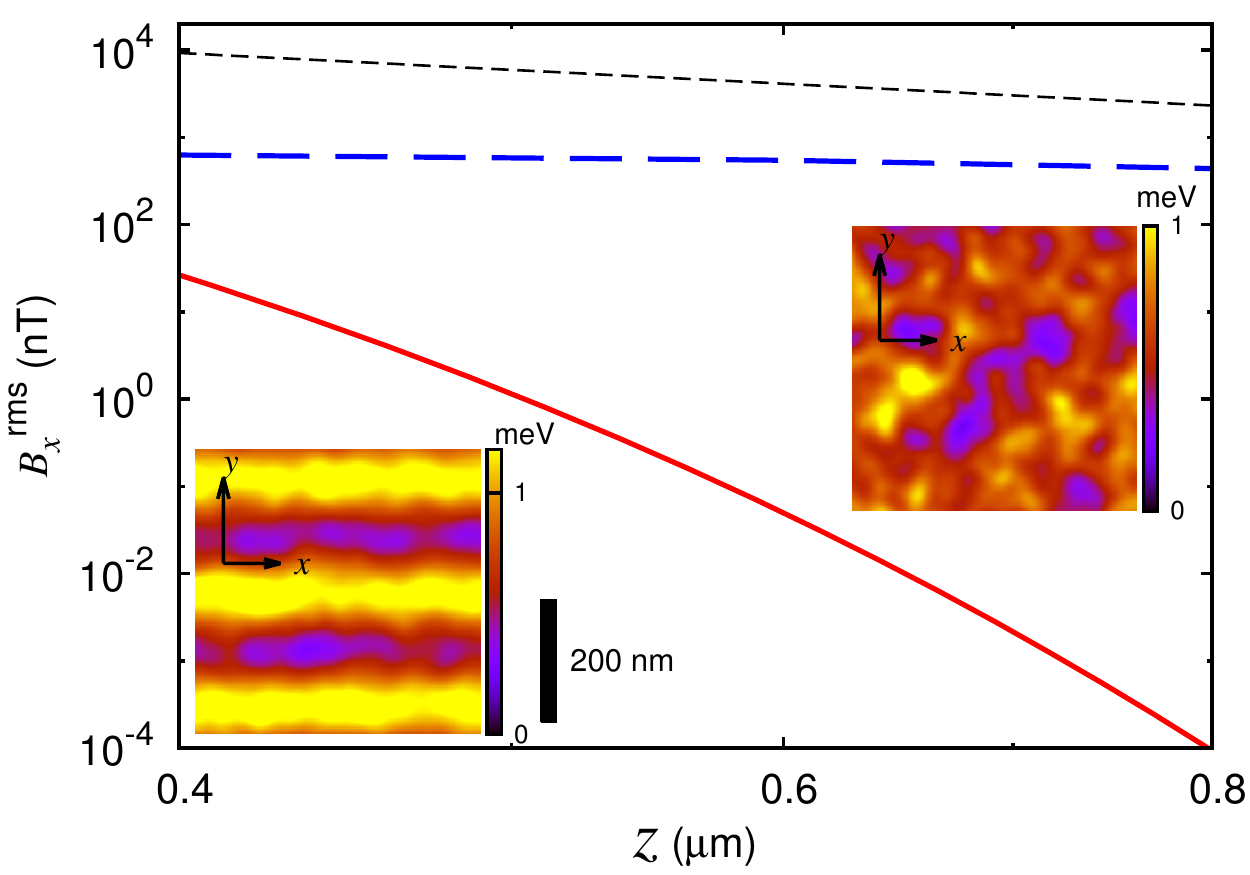} 
  \caption{ \label{B_x_decay_BS}(Colour) $B_{x}^{\text{rms}}(z)$ calculated above a current-carrying 2DEG with spatially random (dashed/blue curve) and periodically-modulated (solid/red curve, modulation period $\lambda=200~$nm) ionised donor profiles in the $x-y$ plane (right- and left-hand insets respectively, which show the potential landscape in the plane of the 2DEG). Black/short-dashed curve: $B_{x}^{\text{rms}}(z)$ produced by edge and surface inhomogeneity above a metal wire of width $W=3~\mu$m carrying a current $I=035~$mA, assuming imperfection with white spectral noise and a typical grain size of $80$ nm \cite{EPJD32_171}. The 2DEG has a mean electron density $n=3.3\times10^{15}~$m$^{-2}$, mobility of $\mu=140~$m$^{2}$V$^{-1}$s$^{-1}$ and is separated from the donor layer by $d=52.9~$nm.}
\end{figure}

As a quantitative example of this control of field inhomogeneity, in figure \ref{B_x_decay_BS} we plot the root-mean-square (rms) amplitude of the magnetic field fluctuations, $B_{x}^{\text{rms}}(z)$, as a function of the distance $z$ normal to the 2DEG plane ($z=0$) with typical experimental parameters. In particular, for these calculations we consider a DC current density $j=100~$Am$^{-1}$ passing through a 2DEG of mean electron density $n=3.3\times10^{15}~$m$^{-2}$ and mobility of $\mu=140~$m$^{2}$V$^{-1}$s$^{-1}$. The dashed blue curve shows $B_{x}^{\text{rms}}$ calculated for an isotropic random distribution of donors, which produces an electric potential landscape, $\Phi(x,y)$, shown in the right inset in figure \ref{B_x_decay_BS}. The solid red curve shows $B^{\text{rms}}_x(z)$ calculated for a patterned 2DEG produced by periodic spatial modulation of the ionised donor density with a period of $200~$nm along the $y$ direction. The corresponding electric potential landscape in the plane of the 2DEG is shown in the left inset of figure \ref{B_x_decay_BS}. For comparison with standard metallic conductors, we also calculate the corresponding $B_x^{\text{rms}}(z)$ curve for field fluctuations produced by edge imperfections in a metal wire (with white spectral noise and grain size of $80$ nm, \cite{rough3}), taking a wire of width $W=3~\mu$m and thickness $t=1~\mu$m, carrying a current $I=0.36~$mA (black/short-dashed curve) \cite{rough3,PRL92_076802}.

As shown in Appendix \ref{sec:field}, near a 2DEG produced by ionised donors with an isotropic spatially random distribution, $B_{x}^{\text{rms}}(z) \propto \frac{1}{(z+d^2)}$ [see Eq. (\ref{B_x_rms_final})], where $d$ is the distance between the 2DEG and the layer of ionised donors. This variation (dashed/blue curve in Fig. \ref{B_x_decay_BS}) lies below the field fluctuations produced by the metal wire (short-dashed/black curve in Fig. \ref{B_x_decay_BS}) for all $z$ in Fig. \ref{B_x_decay_BS} \cite{PhysRevA.76.063621}. Periodic modulation of the 2DEG produces exponential suppression of the field fluctuations, $B_{x}^{\text{rms}}(z) \propto \exp(-2 k_0(z+d))$ where $k_0 = 2\pi/\lambda$ is determined by the modulation period $\lambda$ [see Eq. (\ref{eq:expDecay})]. For the value of $\lambda=200~$nm, corresponding to illumination by an optical standing wave, the field fluctuations (red curve in Fig. \ref{B_x_decay_BS}) are six orders of magnitude lower than for the unpatterned 2DEG at $z = 0.8~\mu$m. Smaller periods of a few $10$s of nm can be produced by electron-beam lithography, making the field fluctuations negligible beyond $100~$nm from the 2DEG. 

The flexibility of modulating the donor distribution allows us to strongly reduce the spatial inhomogeneity of magnetic fields produced by electric currents in 2DEGs. In turn, this enhances the quality of the trapping potential (see section \ref{Sec:trappingandsplitting}) and helps us to reach operational regimes that are inaccessible with standard atom chip platforms, in particular, to produce smooth and tight trapping potential located at submicron distances from the 2DEG.

\section{Trapping and control of BECs with a 2DEG atom chip}
\label{Sec:trappingandsplitting}
We now turn our attention to the use of conducting channels defined in a 2DEG for near-surface trapping and control of ultracold atomic ensembles, as shown in figure \ref{geometry}. We consider conducting channels defined and enclosed by insulating regions (grey in figure \ref{geometry}) made by implanting Ga ions into the heterojunction or by etching it \cite{ion_beam2,ion_beam3}. The general idea is that controlled currents through such structures influence the behaviour of neighbouring ultracold atoms.  Metal contacts deposited on top of the cap layer of the heterostructure provide control over the shape of the channels and current distribution, via the voltages applied to them.

At short atom-surface separations, the nearby surface can compromise the quality and stability of magnetic traps \cite{noise3,EPJD32_171,Henkel1999}. In section \ref{nearsurfacetrapping}, we quantify these effects for the case of a free-standing semiconducting heterojunctions containing a 2DEG.

\subsection{Properties of a magnetic microtrap using a 2DEG conducting channel}
\label{sec:Trap}
To evaluate the ability of 2DEGs to create magnetic traps for ensembles of cold alkali atoms, we consider a single-wire microtrap configuration with a flat conducting channel of width $W$ defined in the 2DEG. For simplicity, we first ignore the effects of atom-surface attraction and evaluate the trapping parameters for typical operating conditions of free-standing heterojunctions.

In our scheme, we consider magnetic trapping of $^{87}$Rb in the state $\left\vert F=2,m_F=2\right\rangle$. The trap comprises the magnetic field produced by a current-carrying Z-shaped channel defined in the 2DEG bounded by the two Z-shaped grey insulating regions fabricated in the 2DEG in Fig. \ref{geometry}, combined with an uniform magnetic field $\boldsymbol{B}_{\text{ext}}= (B_x, B_y, B_z)$, represented by the magenta arrow in Fig. \ref{geometry}. This external field can be produced either by external coils or additional on-chip conductors. The  position of the trap is controlled by the  component of $\boldsymbol{B}_{\text{ext}}$ orthogonal to the central section of the 2DEG Z-shaped channel, which is   $\boldsymbol{B}_{\textrm{\tiny{bias}}} = (0,B_y,B_z)$. The component of $\boldsymbol{B}_{\text{ext}}$ parallel to the mean current flow in the wire [here $\boldsymbol{B}_{\textrm{\tiny{offset}}} = (B_x,0,0)$] provides control over the tightness of the trap and helps to reduce the rate of spin-flip losses (see Sec. \ref{sec:life_time} and \cite{APB74_469}).

In this single-wire microtrap, the intensity and length scale of variations of the magnetic field are set by the current density in the conductor, $\boldsymbol{j}$, and the conductor's width, $W$, respectively. It is convenient to scale the magnetic field by its value at the surface of the conductor, $B_s = \mu_0 |\boldsymbol{j}| / 2$, and define a corresponding energy scale   $E_s := g_F \mu_B  m_F B_s$. A frequency scale (or, equivalently, a time scale) is conveniently defined by $\nu_0 := \frac{1}{2\pi}\sqrt{\frac{E_s}{m W^2}}$, where $m$ is the mass of the trapped atom \cite{SinucoPhDThesis}. 

\begin{figure}[!h]
\centering
\includegraphics[width=12.5cm]{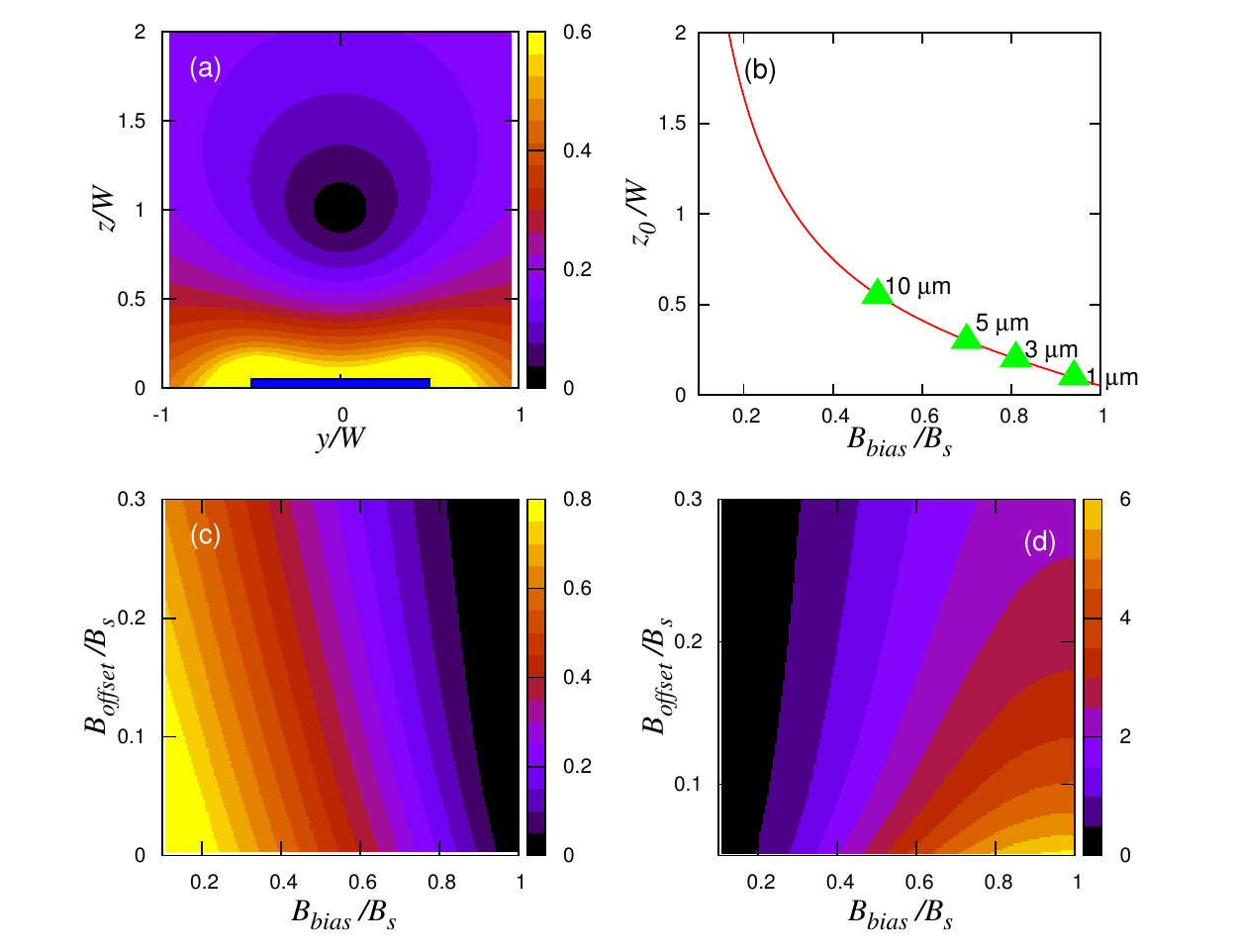}
  \caption{\label{onewire_trapping_potential}Properties of a single-wire magnetic microtrap as functions of the components of an external magnetic field $\boldsymbol{B}_{\textrm{\tiny{ext}}}=\boldsymbol{B}_{\textrm{\tiny{bias}}} + \boldsymbol{B}_{\textrm{\tiny{offset}}}$ (see text). (a) Contour plots of scaled potential energy for the single-wire microtrap configuration.  Here, $B_{\textrm{\tiny{offset}}}=0$ and  the conductor is shown schematically (in cross section) as a blue rectangle. Energy contours are equally spaced by $0.0375 E_s$. Colour scale right is in units of $E_s$.  The bias field is set to produce a potential energy minimum at $z/W \sim 1$. (b) Vertical position  of the trap $z$, scaled to $W$, calculated as a function of the scaled bias field ($B_{\textrm{\tiny{bias}}}/B_s$). Triangles indicate the magnetic field ratio needed to produce traps centred at the indicated distances, with a conductor of width $W=20~\mu$m.  (c)  Contour plots of the scaled trap depth, $\Delta E(B_{\textrm{\tiny{bias}}}/B_s,B_{\textrm{\tiny{offset}}}/B_s)$, as a function of the scaled bias and offset fields. Contours are separated by $0.05E_s$ (colour scale right in units of $E_s$). (d) Contour plot  of the scaled trap frequency along a direction transverse to the plane of the wire, as a function of the scale bias and offset fields. Contours are separated by $0.5\nu_0$ (colour scale right in units of $\nu_0$).} 
\end{figure}

We characterise the quality of a magnetic trap neglecting the effect of gravitational attraction. The height of the energy barrier to the nearest surface defines the trap depth (i.e. the uppermost layer of the heterostructure) and the trap frequency, $\nu_z$, is defined as the curvature of the potential energy at the point of mechanical equibrium along the direction transverse to the plane of the 2DEG \cite{RMP79_235}. These two quantities are functions of the ratios $|\boldsymbol{B}_{\textrm{\tiny{bias}}}|/B_s$  and $|\boldsymbol{B}_{\textrm{\tiny{offset}}}|/B_s$ multiplied by corresponding scaling factors \cite{RMP79_235, SinucoPhDThesis}.  Figure \ref{onewire_trapping_potential} shows the properties of a single-wire micro-trap setup, using the scaling units defined above, which allows us to quickly estimate the values we can obtain in typical atom chip setups. 

Figure \ref{onewire_trapping_potential}(a) shows the potential energy landscape in the $y-z$ plane normal to the central arm of the Z-shaped 2DEG conducting channel (whose cross-section is shown as a blue rectangle at the bottom of the figure), for the case of a bias field adjusted to produce a trap at distance $z=W$ (central black area), where the colour palette indicates the potential energy in units of $E_s$.  Figure \ref{onewire_trapping_potential}(b) shows the variation of the trap minimum relative to the 2DEG plane ($z=0$) versus the magnitude of the applied bias field, $B_{\text{bias}}$, along with a few numerical values for the case of a conductor of width $W=20~\mu$m, indicating the ratios of $B_{\text{bias}}/B_s$ required to locate the trap centre at $z=1~\mu$m, $3~\mu$m, $5~\mu$m and $10~\mu$m. Note that as the bias field increases, the trap centre moves towards the chip surface at $z=0$. Figures \ref{onewire_trapping_potential}(c) and (d), show the trap depth and frequency, respectively, as functions of the scaled bias and offset fields. For a fixed value of the offset field and increasing bias field, the trap depth reduces despite the trap frequency becoming larger. This behaviour results from the displacement of the trap centre towards the 2DEG surface ($z=0$) as the bias field increases. In contrast, for a fixed bias field (or, equivalently, a fixed trap position) the trap depth and frequency both reduce as the offset field increases, which follows from the weakening of the trap.

The depth and frequency of the trap are determined by a combination of geometrical factors (e.g. the shape and dimensions of the conductor) and the strength of the magnetic field produced by the current through the chip. For the single-wire magnetic trap, the intensity of the magnetic field  is limited by the peak current density, $j_{\text{max}}$, supported by the conductor. In addition,  the power dissipated by the elements in the chip should be  small enough to ensure that thermal damage is avoided. This last condition  can be satisfied easily when the conductors operate in a regime of large conductivity, which is one of the reasons why metals have so far been the preferred material for magnetic micro-traps.     

High-mobility 2DEGs reach their peak conductivity at cryogenic temperatures. Typically, they  can sustain currents dissipating a power of a few $10^6~$W m$^{-2}$, corresponding to peak current densities of $j_{\text{max}} \sim  300~$A m$^{-1}$ \cite{APL41_277}.  This value sets the scale of the magnetic fields to  $B_s \simeq 1.89~$G and the energy scale to $E_s \simeq 126.6~\mu$K (for $^{87}$Rb). Using a channel width $W=20~\mu$m, the frequency scale corresponds to  $\nu_0 \simeq 914~$Hz. These simple considerations allow us to identify  sets of parameters that produce magnetic traps with properties similar to those commonly used in magnetic trapping experiments (triangles in Figure \ref{onewire_trapping_potential}(b)) \cite{PhysRevA.76.063621,aigner2008long}, with some specific examples in Table \ref{tab:table1}.

\begin{table}[!h]
\tbl{Set of parameters for a 2DEG-based single-wire magnetic microtrap for cold atoms, comprising a Z-shaped wire of width $W=20~\mu$m defined in a 2DEG with an electron mean density of $n=3.3\times10^{15}~$m$^{2}$ and mobility $\mu=140~$m$^2$V$^{-1}$s$^{-1}$. For all cases, the current density is $j=300~$A/m (corresponding to the total current $I=6~$mA) and the offset field is $B_{\text{offset}}=0.2~$G. As before, $^{87}$Rb in the $\left\vert F=2,m_F=2\right\rangle$ state is considered.}
{
\begin{tabular}{cccc}
\hline
$B_{\text{bias}}$ (G)& $z_0~$($\mu$m)& $\nu_z~$ (kHz) & depth ($\mu$K)\\
\hline
1.54 & 3.0  & 3.27 & 15.2\\
1.33 & 5.0  & 2.85 & 27.9\\
0.94 & 10.0 & 1.78 & 50.6\\
\end{tabular}}
\label{tab:table1} 
\end{table}

Our results indicate that current-carrying conducting channels in 2DEGs can define magnetic traps with spatial frequencies in the kHz range, requiring offset fields of a few hundred mG. Such control of magnetic field strength can be achieved by using chip configurations with a number of different conducting channels, as demonstrated in \cite{PhysRevA.76.063621,EPJD7_361,PRA71_063619}.  Larger trap frequencies and depths will be produced by thinner conductors, making accessible trapping temperatures in the range of $1~\mu$K at the shortest distances. Note also that in this example, the atom chip should be cooled to liquid helium temperatures and thus presents similar challenges to recently-developed superconducting atom chips \cite{PhysRevApplied.7.034026,EPL81_56004, PhysRevLett.98.260407,PhysRevLett.114.113003}.  

\subsection{Casimir-Polder attraction in a magnetic microtrap with a free-standing 2DEG}
\label{nearsurfacetrapping}
The above discussion suggests that 2DEGs can create near-surface magnetic microtraps with properties that allow coupling between atomic degrees of freedom and quantum electronic devices fabricated within the chip. Long lifetimes of the trapped atomic states are expected and the roughness of the magnetic field  produced by the 2DEG can be reduced by periodic modulation of the ionised donor distribution. This allows us to  prepare strong and smooth trapping configurations to store atoms at submicron distances from the atom chip (see figures \ref{B_x_decay_BS} and \ref{onewire_trapping_potential}), where the atoms can directly couple to charge carriers of semiconducting devices \cite{PhysRevA.83.021401,0953-4075-46-24-245502}.

At submicron atom-surface distances, the Casimir-Polder (CP) attraction between the surface and the atoms can no longer be ignored \cite{noise3}. CP attraction, however, should be weak for suspended semiconductor membranes of thickness  $\le 10~\mu$m, such as ultrathin heterostructures containing a 2DEG, ultrathin SiN and graphene sheets. 

\begin{figure}[!h]
  \centering
  \includegraphics[width=7.5cm]{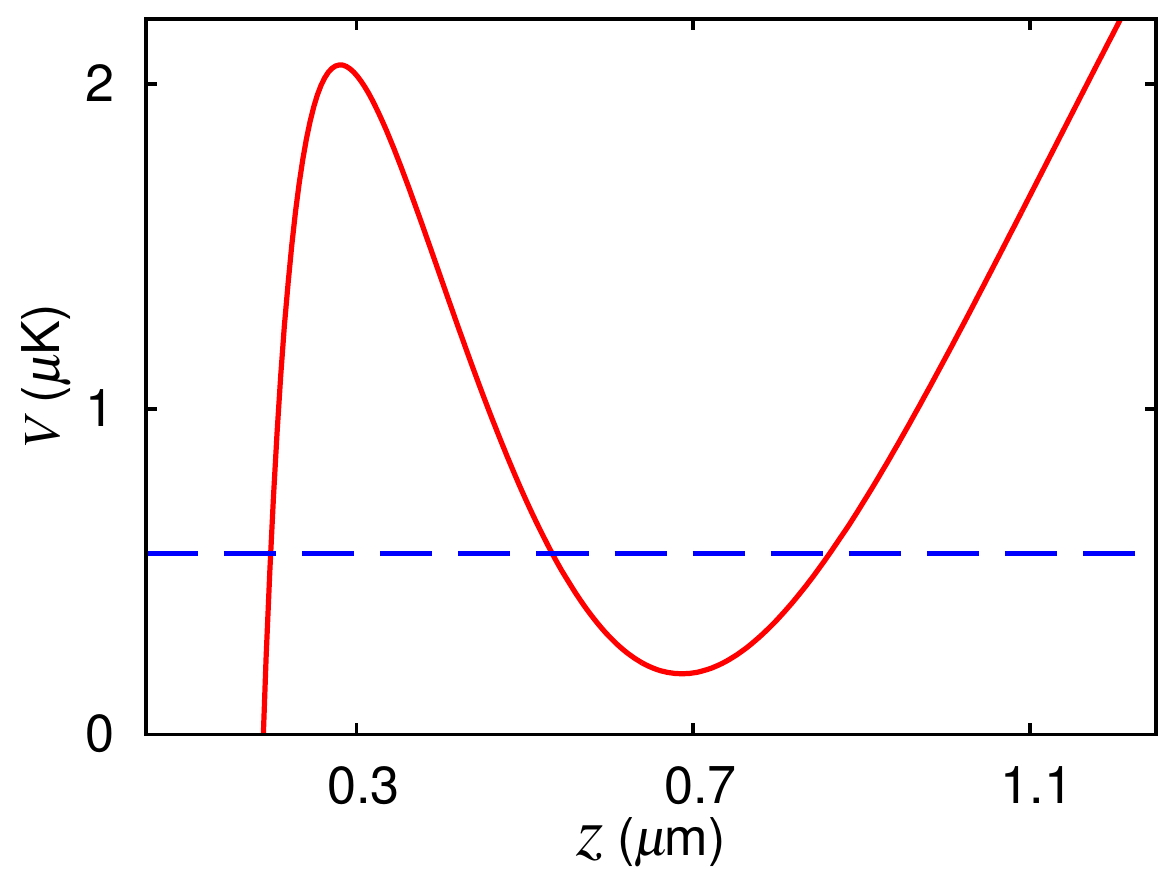}
  \caption{Total potential energy, $V=V_m+V_{\text{CP}}$, (red/solid curve) calculated as a function of $z$ for the ground state $\left|F=2,m_F=2\right\rangle$ of $^{87}$Rb. Blue/dashed line: chemical potential of a trapped BEC with  $500$ atoms of $^{87}$Rb in the state $\left|F=2,m_F=2\right\rangle$. The magnetic potential, $V_m$, is created by combining the magnetic field produced by a Z-shaped conductor of width $W=3~\mu$m carrying a current density of $j=118~$A m$^{-1}$ (i.e. $I=0.35~$mA) with an external homogeneous field $\boldsymbol{B}_{\text{ext}}=(40,536,0)~$mG. The Casimir-Polder contribution, $V_{\text{CP}}$, is calculated following \cite{PhysRevA.80.032901}.}
  \label{QPC_BEC_potential} 
\end{figure}

The attractive CP potential, $V_{CP}(z)$, can be calculated using Eqs. (25-29) of Ref. \cite{PhysRevA.80.032901}, which are valid at any vertical distance $z$ from a uniform dielectric slab. For the present case, $V_{CP}(z)$ is determined by the coefficient $C_4  = 2 \times 10^{-54}$ Jm$^4$ for Rb atoms near GaAs \cite{PhysRevA.80.032901}.  To  evaluate the impact of this strongly attractive potential on the quality of the 2DEG-based trap described in section \ref{sec:Trap}, we consider a Z-shaped 2DEG channel of width $3~\mu$m and central arm length of $60~\mu$m, carrying a current density $j=118~$A m$^{-1}$, corresponding to a current $I=0.35~$mA. The solid curve in Fig.~\ref{QPC_BEC_potential}(a) shows the total potential energy $V(z)=V_m(z)+V_{CP}(z)$ calculated for $^{87}$Rb atoms in the state $\left|F=2,m_F=2\right\rangle$ of the ground state manifold, where $V_{m}(z)$ originates from the magnetic field produced by the current through the 2DEG channel and an applied field $\boldsymbol{B}_{\text{ext}}=(40, 536, 0)$ mG.  In this case, the CP attraction is overcome by the magnetic potential and the trap is deep enough to confine a small ultracold ensemble of atoms, for example a Bose-Einstein condensate comprising $500$ $^{87}$Rb atoms in the hyperfine state $\left|F=2,m_F=2 \right\rangle$, whose chemical potential (horizontal line in Fig.~\ref{QPC_BEC_potential}) is far below the top of the energy barrier nearest the chip surface.

\subsection{Effect of quantised electronic atom chip conductance on the density profile of a nearby BEC}
\begin{figure}[!h]
  \centering
  \includegraphics[width=8.5cm]{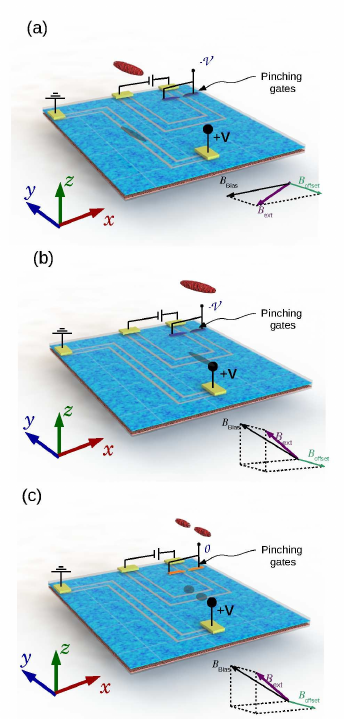}
  \caption{Sequence for controlled splitting of a BEC via opening a quantised conductance channel in a U-shaped 2DEG conductor (towards the ar right region of the blue plane), bounded by the grey insulating lines fabricated in the 2DEG. The conductance of the U channel is controlled by a negative voltage applied to two pinching gates (blue/orange), one located at each side of one of the channel arms. (a) A BEC (red) is trapped by a magnetic trap made by a current-carrying Z shaped 2DEG conductor defined on a 2DEG and an external magnetic field (magenta arrow) that provides the bias and offset fields. (b) The  BEC is moved above the middle arm of a U-shaped conductor by adjusting the current in the Z channel (via the voltage \textbf{V}) and tilting the  external magnetic field. The vertical position of the trap centre is adjusted to $z=0.7~\mu$m. The voltage applied to the two surface pinching gates (blue patches near the upper-right corner of the 2DEG plane) is negative enough to depopulate all conduction channels, preventing current through the U-shaped channel. (c) Reducing the magnitude of the negative voltage applied to the pinching gates (i.e. $-\mathcal{V}<0 \rightarrow -\mathcal{V}=0$, as indicated by changing the colour of the pinching gates from blue to orange), one quantised conduction channel in the U conductor is opened to let current flow. The small local magnetic field created by this current splits the BEC. The shadow of the BEC over the chip surface is added to guide the eye.}
  \label{QPC_BEC_procedure} 
\end{figure}

Near-surface trapping makes the atomic gas highly sensitive to magnetic field variations arising from the geometry of the conducting channels, including local narrowing. As an example, suppose that the magnetic fields of the trapping configuration explained above are adjusted to place the BEC across the middle arm of a U-shaped channel, fabricated next to the Z-shaped trapping channel, as shown schematically in Fig. \ref{QPC_BEC_procedure}. 

When a small current is passed through the U-shaped conductor ($I_{\text{QPC}}\approx 5~\mu$A), the potential energy of the atoms in the BEC rises  directly above the channel.  The corresponding reduction of the BEC's local density therefore sensitively reflects the conducting state of the channel, even to the level of registering discrete changes of the channel conductance (see Fig. \ref{QPC_BEC_procedure}). Quantised steps in the channel conductance can be swept through by changing a negative voltage applied to metal surface gates ((blue/orange) in Fig. \ref{QPC_BEC_procedure}) positioned either side of the arms of the U-shaped conductor in the 2DEG. This negative voltage will produce a local narrowing of the arm, which can support an integer number of propagating electronic modes contributing to the channel transport \cite{Bastard}. As the voltage is made more negative, the number of propagating modes decreases until the last one is depopulated and the conductor channel is pinched off. 

 \begin{figure}[!h]
    \centering   
   \includegraphics[width=12.5cm]{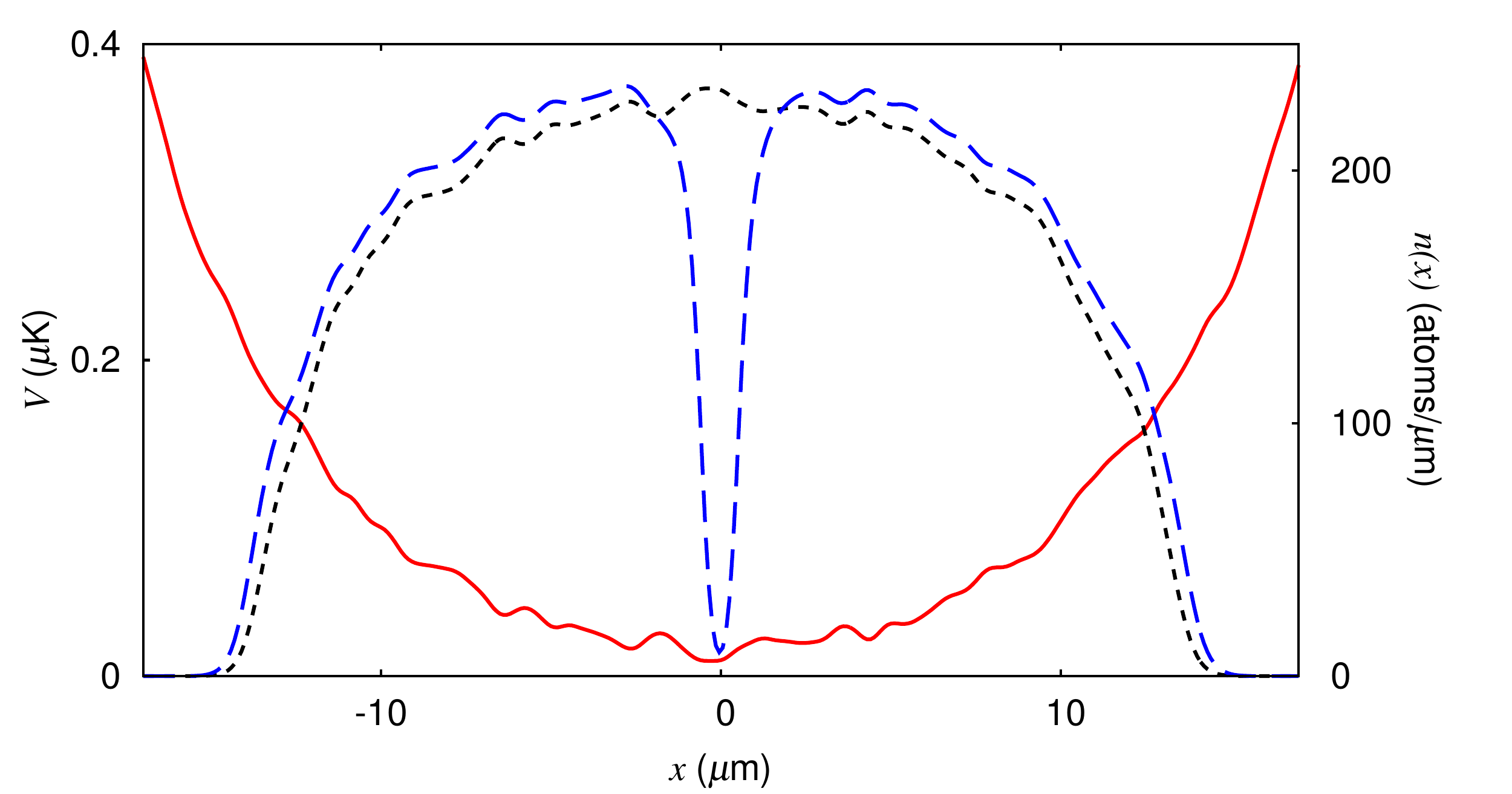}
   \caption{Red/solid curve: Calculated total potential energy along the $x-$axis of a quasi 1D BEC trapped  at $z=0.7~\mu$m above the centre of the Z-shaped 2DEG conductor (Fig.~\ref{QPC_BEC_procedure}). The inhomogeneity reflects a residual meandering of the currents in the 2DEG with a periodic pattern of period $200~$nm. Short-dashed (black) and dashe (blue) curves: atom density profiles, $n(x)$ of the BEC above the U-shaped conductor in Fig. \ref{QPC_BEC_procedure} with $N=0$ and $N=1$ open quantised conductance channels, respectively.}
      \label{QPC_BEC_density} 
 \end{figure}

In figure \ref{QPC_BEC_density} we present a quantitative example of this splitting mechanism, comparing the 1D density profiles of the atom cloud for two distinct conducting states of a U-shaped channel. The trapping potential has a residual inhomogeneity due to the meandering of the currents in the 2DEG, which are strongly suppressed by assuming a periodic pattern of period $200~$nm (see Section \ref{sec:fieldinhomegneities}). The short-dashed/black curve in figure \ref{QPC_BEC_density} shows the 1D density profile\footnote{The 1D density is produced by integrating the 3D density over  the transverse cross-section.}, $n(x)$, of the BEC when the number of open channels in the depletion region of the U-shaped channel is $N=0$. Since there cannot be current flow in this case, $n(x)$ is just the unperturbed ground state density profile of the trapping potential shown by the solid/red curve in figure \ref{QPC_BEC_density}. Opening \emph{a single} quantised conduction channel ($N=1$) in the U-shaped conductor changes the trap profile sufficiently to almost completely split the BEC [dashed/blue curve in figure \ref{QPC_BEC_density}]. Thus, even the smallest quantised changes of the conductance can either be detected by using the BEC or actively used for manipulating the atom density profile.
\clearpage

\section{BEC magnetometry of a 2DEG}
\label{sec:magnetometry}
The transport properties of semiconductor devices depend strongly on the spatial distribution of their constituent materials at both long-range and  atomic scales. As extreme situations,  we have fully ballistic transport for perfect crystalline structures and, in contrast, diffusive transport in media with high defect density. In recent years, by bringing ensembles of alkali atoms close to microfabricated electronic devices, BEC magnetometry \cite{Nature_wildermuth} and microwave atomic scanning  \cite{1367-2630-17-11-112002,bohi2010imaging} have opened opportunities for investigating electron transport phenomena with unprecedented spatio-temporal resolution. Thanks to the sensitivity of the atom cloud's dynamics to external fields, and modern high-precision knowledge of the atomic structure, these developments make it possible to relate spatial inhomogeneity in the optical images of atomic ensembles directly to the electronic properties of the solid-state device under study \cite{bohi2010imaging, Nature_wildermuth,aigner2008long}. 

\begin{figure}[!h]
  \centering 
  \includegraphics[width=9.5cm]{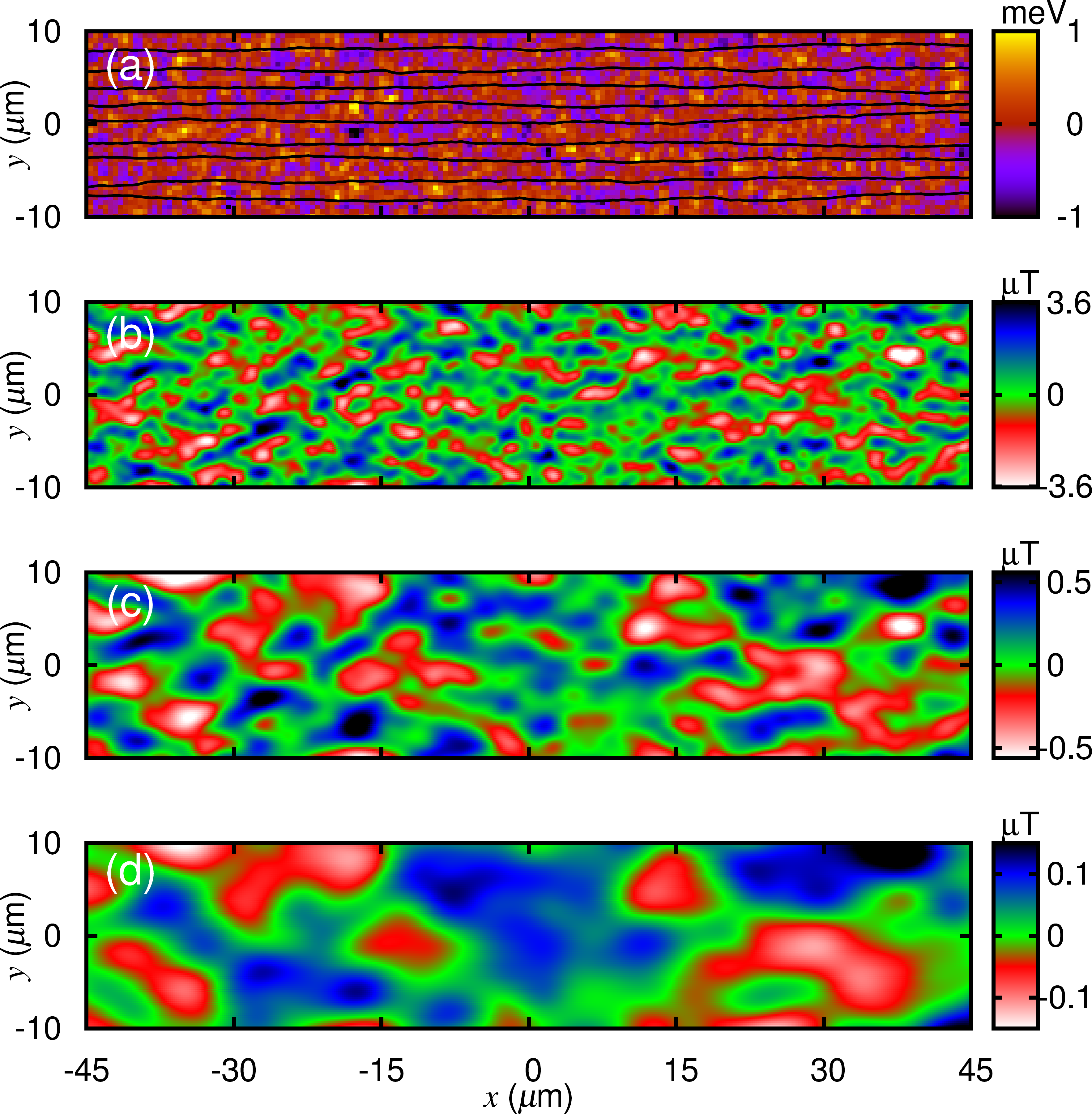}
  \caption{(Colour) (a) Calculated electrostatic potential energy landscape of electrons in a 2DEG, $\Phi(x,y)$. Black curves show current stream lines calculated for an electric field of $E = 1.6~$mV/$\mu$m applied along the conducting channel, which produces a mean current density $j=118~$Am$^{-1}$ in a 2DEG of mean electron density $n=3.3\times10^{15}~$m$^{-2}$ and mobility $\mu=140~$m$^{-2}$V$^{-1}$s$^{-1}$. The other panels show the $x$ component of the magnetic field calculated above the 2DEG in the $z=~$ (b) $1~\mu$m, (c) $3~\mu$m  and (d) $5~\mu$m planes. \label{two_deg_properties_fig_and_B_x_at}. For each panel, color scales are shown right.}
\end{figure}

In this section, we consider what BEC magnetometry can tell us about the structure of a semiconductor heterostructure containing a 2DEG. In such devices, the electron mobility is influenced by the spatial distribution of the dopants that provide the charges carriers (here electrons) in the 2DEG. In typical GaAs/(AlGa)As heterostructures, Si donors are confined to a thin layer ($\delta$-doping) located at $z=-52.9~$nm from the 2DEG plane. The ionised Si atoms in the heterostructure $\delta-$doping layer create an inhomogeneous electrostatic potential landscape for electrons in the 2DEG [fig. \ref{two_deg_properties_fig_and_B_x_at}(a)], and, thereby limit its transport quality \cite{High_mob}.  Generally, the statistical properties of the donor distribution are hard to measure directly without strongly perturbing the device (e.g. using a scanning tunnelling microscope \cite{Fleischmann}). However, BEC magnetometry offers to overcome this challenge by mapping the inhomogeneity of the magnetic field created when a small electric current passes through the 2DEG  \cite{Buks2,Coleridge,Grill,xray,Fleischmann}. Moreover, the magnetic field profile provides direct information about the potential energy landscape in the 2DEG plane and of the underlying ionised donor distribution.

In the 2DEG, the inhomogeneous electronic potential energy landscape, $\Phi(x,y)$, created by the ionised donors, disturbs the rectilinear trajectories that would follow under the action of a uniform electric field. Typical current stream lines are shown in figure  \ref{two_deg_properties_fig_and_B_x_at}(a), calculated for a uniform electric field of $1.6 \times 10^{3}~$ V m$^{-1}$ applied along the $x$ direction in the plane of the 2DEG. The features of the magnetic field fluctuations created by this current smooth out as the distance from the 2DEG plane increases, as shown in figures  \ref{two_deg_properties_fig_and_B_x_at}(b)-(c) \cite{PhysRevA.83.021401}. With BEC magnetometry, such field fluctuations can be mapped by scanning the $y$ position of a quasi-1D BEC stretched along the $x$ axis, and, for each $y$ value,  measuring its density , $n(x)$, along the $x$-direction \cite{Nature_wildermuth}.

To quantify this idea, we calculate the atomic density of a needle-like atomic BEC confined in a magnetic trap with trapping angular frequencies in the ratio $\omega_\bot : \omega_{x} = 100:1$, where $\omega_\bot$ ($\omega_x$) is the radial (axial) trapping frequency in the $y-z$ plane ($x$ direction). The BEC comprises $10^4$ $^{87}$Rb atoms in hyperfine ground Zeeman state $\left\vert F=2,m_F=2\right\rangle$ and is positioned at several distances, $z$, from a current-carrying 2DEG (see figure \ref{geometry}). The BEC density modulations mirror the magnetic field fluctuations created by the current in the 2DEG [Figs. \ref{two_deg_properties_fig_and_B_x_at}(b)-(d)]. By confining the BEC strongly along the $y$ and $z$ directions, the  atom density profiles are sensitive only to fluctuations in the field component $B_x(x,y,z)$,  produced  by the $y$-component of the current in the 2DEG.  Within the Thomas-Fermi approximation, along the length of the BEC, which is parallel to the  $x-$axis, the magnetic field and atom density fluctuations are related by \cite{APL88_264103}: 
\begin{equation}
B_x(x,0,z) = -2 \hbar \omega_\bot a_s \Delta n(x,0,z)/ (m_F g_F \mu_B),
\label{densityBECandField}
\end{equation}
where  $a_s$ is the s-wave atomic scattering length, $g_F$ is the Land\'e g-factor and $\mu_B$ is the Bohr magneton.

\begin{figure}[!htb]
\centering
\includegraphics[width=14cm]{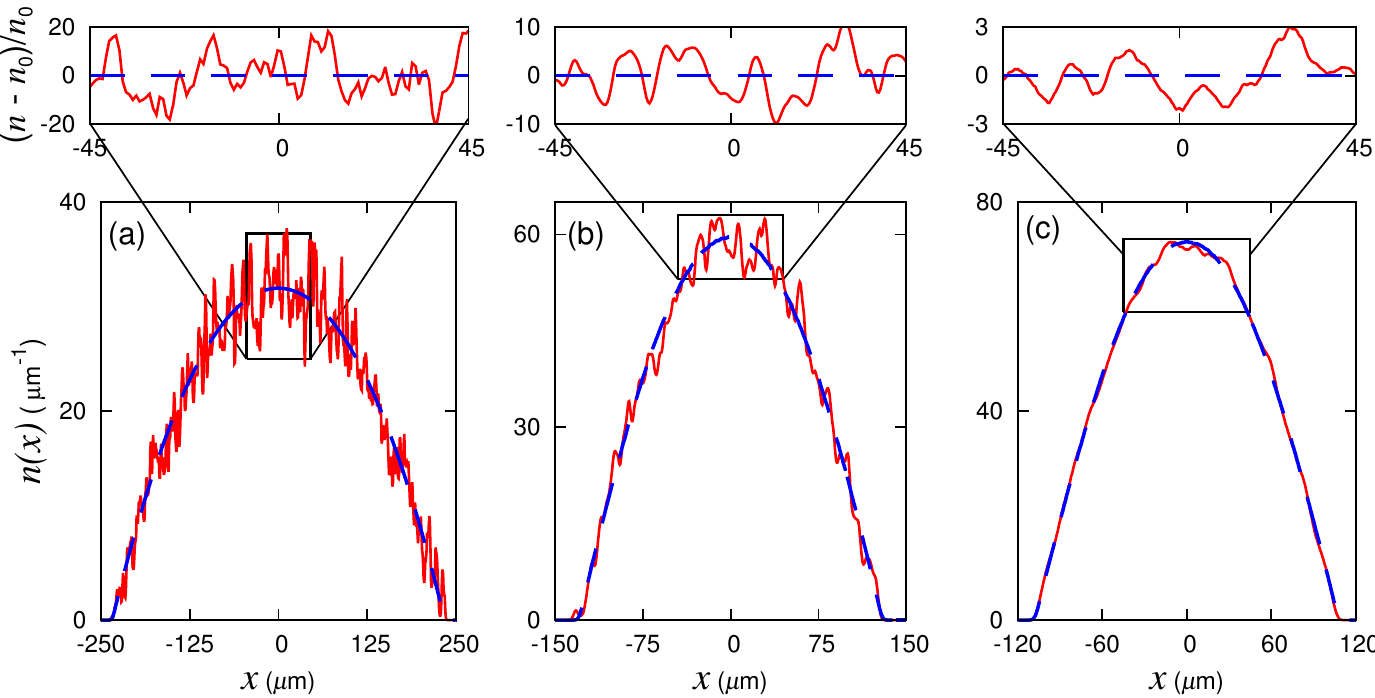}
\caption{\label{fig:gs1d_2deg_trap}
Atom density profiles, $n(x)$, calculated along the direction of the trapping conductor, at distances (a) $z=1~\mu$m, (b)   $z=3~\mu m$ and (c) $z=5~\mu$m above a 2DEG with (solid/red curves) and without (dashed/blue curves) current flow in the 2DEG. Upper panels: relative fluctuations of the atom density profiles along the centre of the BEC, showing the fluctuations (red curve) produced when current flows through the 2DEG.}
\end{figure}

Figure \ref{fig:gs1d_2deg_trap} shows the 1D density profile of an elongated BEC trapped at $z_0= 1~\mu$m, $3~\mu$m and $5~\mu$m from  a current-carrying 2DEG. Insets (a)-(c), respectively, show enlargements of the central region and variations of the atom density relative to its value in a trap without inhomogeneity. Typical density modulations are $\approx 20~$\% of the atom density at $z=1~\mu$m [Fig. \ref{fig:gs1d_2deg_trap}(a)], falling below the present detection limit ($\approx 10~$\%) at $z\approx 5~\mu$m [Fig. \ref{fig:gs1d_2deg_trap}(c)] \cite{PhysRevA.76.063621}. 

Since the current density is confined to a 2D plane, the $x$-component of the magnetic field component, $B_x$, and the $y$-component of the current density component, $j_y$, have a simple relation in terms of Fourier transforms  \cite{JAP65_361,APL88_264103}:
\begin{equation}
B_x(x,y,z_0) =  \frac{\mu_0}{4} \int k_y e^{-k z_0} e^{i(k_x x + k_y y)} \left( \frac{1}{4 \pi^2} \int  j_y(x',y') e^{-i(k_x x' +k_y y')}dx'dy'\right)  dk_xdk_y,
\label{eq:FieldCurrentFourierRelation}
\end{equation}
where $k=(k_x^2+k_y^2)^{1/2}$. This relation enables us to reconstruct the total current distribution in the 2DEG plane by deconvolution of the atomic density profiles measured at different positions in the $z$ plane \cite{Nature_wildermuth}.  Note that, due to the exponential suppression of the spatial frequency components of $j_y$ in Eq. (\ref{eq:FieldCurrentFourierRelation}), $z$ becomes the maximal spatial resolution for which the current density can be calculated accurately from measurements of the atomic density.

\begin{figure}[!htb]
\centering
\includegraphics[width=14cm]{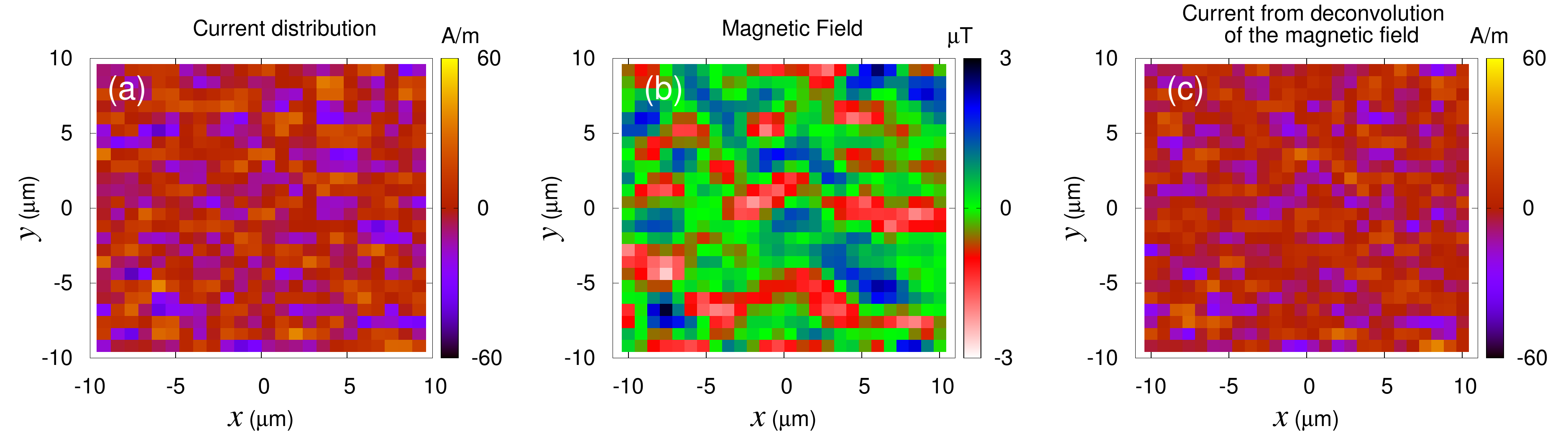}
 \caption{\label{fig:deconvolution} Reconstruction of current density profiles in 2DEG using BEC magnetometry. (a) Low-resolution image calculate for the current density in a 2DEG,  (b) Simulation of the resulting magnetic field profile in the plane $z=1~\mu$m measured by BEC magnetometry (c) current density profile reconstructed from deconvolution of the magnetic field distribution in panel (b).}
\end{figure}

We reconstructed the current density profile by inverting Eq. (\ref{eq:FieldCurrentFourierRelation}) (i.e. performing a numerical deconvolution) for the magnetic field landscape at a distance of $z = 1~\mu$m. To simulate an experimental run, we proceed as follows: we calculate the current profile over an area of $200~\mu$m$\times200~\mu$m of the 2DEG with a spatial resolution of $0.1~\mu$m. Using the Biot-Savart law, this current distribution is used to calculate the magnetic field over an area of $20~\mu$m$\times20~\mu$m with a resolution of $0.8~\mu$m in the plane $z=1~\mu$m, shown in Fig. \ref{fig:deconvolution}(b). This magnetic field profile simulates data obtained by BEC microscopy. The current distribution is then reconstructed by inverting Eq. (\ref{eq:FieldCurrentFourierRelation}) using the calculate magnetic field profile.  The reconstructed current profile in Fig. \ref{fig:deconvolution}(c) should be compared with a low-resolution image of the original current distribution in Fig. \ref{fig:deconvolution}(a). 

Note that features with characteristic lengths larger than $z \approx 1~\mu$m appear in both panels (a) and (c) in Fig. \ref{fig:deconvolution}. The Pearson correlation coefficient between the data shown in panels (a) and (c) is  \textbf{corr}$= 0.7$, indicating a high degree of (at least) linear dependence between the two panels \cite{doi:10.1080/00031305.1988.10475524}. The same procedure applied to the magnetic field distribution at $z = 3~\mu$m and $5~\mu$m produces current distributions (not shown) that resemble the original distribution with a coarser resolution, due to exponential decay of high-spatial frequency components of the magnetic field  in Eq. (\ref{eq:FieldCurrentFourierRelation}) \cite{APL88_264103}.

The electric potential landscape in the 2DEG, $\Phi(x,y)$, can be calculated combining the reconstructed current with the charge conservation law. Since the distribution of ionized donors located below the 2DEG, $n(x,y,-d)$, determines $\Phi(x,y)$ (see Appendix \ref{sec:2DEG}), detection of magnetic field profiles enables us to calculate $n(x,y,-d)$,  avoiding the strong local perturbations of the 2DEG used by other techniques \cite{JAP65_361,Fleischmann,aigner2008long}. 

\section{Summary}
\label{sec:summary}
We have presented a theoretical analysis of atom chip configurations attainable by using free-standing 2DEG heterojunctions. In particular, we considered two complementary aspects: trapping of atomic clouds by the atom chip and, conversely, the use of the atoms to probe the structure and function of the chip.

We have quantified the advantages of using  2DEGs, rather than metals, as the trapping conductors in atom chips, specifically reduced atom loss rates, the ability to tailor the magnetic field inhomogeneity by manipulating the donor distribution, and weak atom-surface attraction. All of these advantages will help to achieve coupling between quantum electronic devices and the centre of mass and internal state degrees of freedom of near-surface trapped atoms. Fabrication advantages can also be envisaged since additional quantum electronic devices, such as quantum dots, can be incorporated in the atom chip within the same production process.

Once the limiting factors to reduce the atom-surface distance are overcome, tiny currents in the chip can produce significant changes in the atom cloud's density. This paves the way to developing more complex applications where solid-state devices are coupled to trapped atoms in their neighbourhood \cite{PhysRevB.92.245439}.  For example, single-electron transistors (SETs) may be switched by the presence/absence  of an atom. Excited atoms may couple to electrons in 2DEGs to modify their dynamics, and atomic Rydberg states may trigger Coulomb blockade in SETs.

Our work also suggests that current advances in BEC magnetometry techniques can provide new insights into the structural and functional properties of semiconductor devices and so contribute to improving their transport properties \cite{Buks2,Coleridge,Grill}. At the moment, BEC microscopy can probe long-range structural features, which strongly affect electron transport in ultra-high mobility 2DEGs \cite{Mico,High_mob,siegert}, and produce single-shot snapshots of time-dependent donor distributions. We foresee that improvements in the resolution and sensitivity of this technique may allow the direct visualisation of many other static and dynamical phenomena including Anderson localisation, conductor-insulating transitions, and electron Wigner-crystals. Non-invasive BEC microscopy has recently revealed striking long-range patterns in the classical current flow through metals \cite{aigner2008long} and may yield similar surprises in other materials including, e.g.,  spin transport in ferromagnetic semiconductors, with the advantage of leaving the system under scrutiny unperturbed \cite{1107.2268}. 

\section*{Acknowledgements}
The authors would like to thank Joe Shearring for helpful discussions. GS would like to thank Barry Garraway for support during the writing of this document, and to Nazri Abdul Halif (Univ. Malaysia Perlis) for enlightening discussions.  This work was funded by EPSRC through the UK Quantum Technologies Hub for Sensors and Metrology  (grant reference EP/I010394/1).

\bibliographystyle{tfp}
\bibliography{extended_B_paper}

\begin{thebibliography}{10}
\providecommand{\url}[1]{\normalfont{#1}}
\providecommand{\urlprefix}{}

\bibitem{FifteenYearsFolman}
Keil, M.; Amit, O.; Zhou, S.; Groswasser, D.; Japha, Y.; Folman, R. Fifteen
  years of cold matter on the atom chip: promise, realizations, and prospects,
  \emph{J. Mod. Opt.}  \textbf{2016}, \emph{63}~(18), 1840--1885.

\bibitem{reichel1999atomic}
Reichel, J.; H{\"a}nsel, W.; H{\"a}nsch, T. Atomic micromanipulation with
  magnetic surface traps, \emph{Phys. Rev. Lett.}  \textbf{1999},
  \emph{83}~(17), 3398.

\bibitem{DeMotte2016}
De~Motte, D.; Grounds, A.R.; Reh{\'a}k, M.; Rodriguez~Blanco, A.; Lekitsch, B.;
  Giri, G.S.; Neilinger, P.; Oelsner, G.; Il'ichev, E.; Grajcar, M.; et~al.
  Experimental system design for the integration of trapped-ion and
  superconducting qubit systems, \emph{Quantum Information Processing}
  \textbf{2016}, \emph{15}~(12), 5385--5414.

\bibitem{cridland2016single}
Cridland, A.; Lacy, J.; Pinder, J.; Verd{\'u}, J. Single microwave photon
  detection with a trapped electron, In \emph{Photonics}, Multidisciplinary
  Digital Publishing Institute, 2016; p~59.

\bibitem{PhysRevLett.100.153003}
Meek, S.A.; Bethlem, H.L.; Conrad, H.; Meijer, G. Trapping Molecules on a Chip
  in Traveling Potential Wells, \emph{Phys. Rev. Lett.}  \textbf{2008},
  \emph{100}, 153003.

\bibitem{PhysRevLett.90.173001}
Hammes, M.; Rychtarik, D.; Engeser, B.; N\"agerl, H.C.; Grimm, R.
  {Evanescent-Wave Trapping and Evaporative Cooling of an Atomic Gas at the
  Crossover to Two Dimensions}, \emph{Phys. Rev. Lett.}  \textbf{2003},
  \emph{90}~(17), 173001.

\bibitem{PhysRevLett.104.203603}
Vetsch, E.; Reitz, D.; Sagu\'e, G.; Schmidt, R.; Dawkins, S.T.; Rauschenbeutel,
  A. Optical Interface Created by Laser-Cooled Atoms Trapped in the Evanescent
  Field Surrounding an Optical Nanofiber, \emph{Phys. Rev. Lett.}
  \textbf{2010}, \emph{104}, 203603.
  \urlprefix\url{https://link.aps.org/doi/10.1103/PhysRevLett.104.203603}.

\bibitem{noise1}
Henkel, C. Magnetostatic field noise near metallic surfaces, \emph{Eur. Phys.
  J. D}  \textbf{2005}, \emph{35}~(1), 59--67.

\bibitem{noise2}
Sinclair, C.D.J.; Curtis, E.A.; Garcia, I.L.; Retter, J.A.; Hall, B.V.;
  Eriksson, S.; Sauer, B.E.; Hinds, E.A. Bose--{E}instein condensation on a
  permanent-magnet atom chip, \emph{Phys. Rev. A}  \textbf{2005}, \emph{72},
  031603.

\bibitem{noise3}
Lin, Y.J.; Teper, I.; Chin, C.; Vuleti\ifmmode~\acute{c}\else \'{c}\fi{}, V.
  Impact of the Casimir-Polder Potential and Johnson Noise on Bose-Einstein
  Condensate Stability Near Surfaces, \emph{Phys. Rev. Lett.}  \textbf{2004},
  \emph{92}, 050404.

\bibitem{zhang2005relevance}
Zhang, B.; Henkel, C.; Haller, E.; Wildermuth, S.; Hofferberth, S.; Kr{\"u}ger,
  P.; Schmiedmayer, J. Relevance of sub-surface chip layers for the lifetime of
  magnetically trapped atoms, \emph{The European Physical Journal D-Atomic,
  Molecular, Optical and Plasma Physics}  \textbf{2005}, \emph{35}~(1),
  97--104.

\bibitem{PRL92_076802}
Wang, D.W.; Lukin, M.D.; Demler, E. Disordered Bose-Einstein Condensates in
  Quasi-One-Dimensional Magnetic Microtraps, \emph{Phys. Rev. Lett.}
  \textbf{2004}, \emph{92}, 076802.

\bibitem{PhysRevA.76.063621}
Kr\"uger, P.; Andersson, L.M.; Wildermuth, S.; Hofferberth, S.; Haller, E.;
  Aigner, S.; Groth, S.; Bar-Joseph, I.; Schmiedmayer, J. Potential roughness
  near lithographically fabricated atom chips, \emph{Phys. Rev. A}
  \textbf{2007}, \emph{76}, 063621.
  \urlprefix\url{https://link.aps.org/doi/10.1103/PhysRevA.76.063621}.

\bibitem{rough1}
Fort\'agh, J.; Ott, H.; Kraft, S.; G\"unther, A.; Zimmermann, C. Surface
  effects in magnetic microtraps, \emph{Phys. Rev. A}  \textbf{2002},
  \emph{66}, 041604.

\bibitem{rough2}
Jones, M.P.A.; Vale, C.J.; Sahagun, D.; Hall, B.V.; Hinds, E.A. Spin Coupling
  between Cold Atoms and the Thermal Fluctuations of a Metal Surface,
  \emph{Phys. Rev. Lett.}  \textbf{2003}, \emph{91}, 080401.

\bibitem{rough3}
Est\'eve, J.; Aussibal, C.; Schumm, T.; Figl, C.; Mailly, D.; Bouchoule, I.;
  Westbrook, C.I.; Aspect, A. Role of wire imperfections in micromagnetic traps
  for atoms, \emph{Phys. Rev. A}  \textbf{2004}, \emph{70}, 043629.

\bibitem{Folman_sub}
Salem~\textit{et al.}, R. Nanowire atomchip traps for sub-micron atom-surface
  distances, \emph{New J. Phys.}  \textbf{2010}, \emph{12}, 023039.

\bibitem{doi:10.1063/1.2219397}
Allwood, D.A.; Schrefl, T.; Hrkac, G.; Hughes, I.G.; Adams, C.S. Mobile atom
  traps using magnetic nanowires, \emph{Appl. Phys. Lett.}  \textbf{2006},
  \emph{89}~(1), 014102.

\bibitem{PRL98_263201}
Trebbia, J.B.; Garrido~Alzar, C.L.; Cornelussen, R.; Westbrook, C.I.;
  Bouchoule, I. Roughness Suppression via Rapid Current Modulation on an Atom
  Chip, \emph{Phys. Rev. Lett.}  \textbf{2007}, \emph{98}, 263201.

\bibitem{APL88_264103}
Wildermuth, S.; Hofferberth, S.; Lesanovsky, I.; Groth, S.; Kr\"uger, P.;
  Schmiedmayer, J.; Bar-Joseph, I. {Sensing electric and magnetic fields with
  Bose-Einstein condensates}, \emph{Appl. Phys. Lett.}  \textbf{2006},
  \emph{88}, 264103.

\bibitem{Nature_wildermuth}
Wildermuth, S.; Hofferberth, S.; Lesanovsky, I.; Haller, E.; Andersson, L.;
  Groth, S.; Bar-Joseph, I.; Kr\"uger, P.; Schmiedmayer, J. Bose--{E}instein
  condensates: Microscopic magnetic-field imaging, \emph{Nature}
  \textbf{2005}, \emph{435}~(7041), 440.

\bibitem{PhysRevApplied.7.034026}
Yang, F.; Koll\'ar, A.J.; Taylor, S.F.; Turner, R.W.; Lev, B.L. {Scanning
  Quantum Cryogenic Atom Microscope}, \emph{Phys. Rev. Applied}  \textbf{2017},
  \emph{7}, 034026.

\bibitem{JPCS19_56}
Kr\"uger, P.; Wildermuth, S.; Hofferberth, S.; Mauritz, A.L.; Groth, S.;
  Bar-Joseph, I.; Schmiedmayer, J. Cold atoms close to surfaces: measuring
  magnetic field roughness and disorder potentials, \emph{J. Phys.: Conf. Ser.}
   \textbf{2005}, \emph{19}, 56.

\bibitem{JAP65_361}
Roth, B.; Sepulveda, N.; Wikswo, J. Using a magnetometer to image a
  two-dimensional current distribution, \emph{J. Appl. Phys.}  \textbf{1989},
  \emph{65}, 361.

\bibitem{aigner2008long}
Aigner, S.; Della~Pietra, L.; Japha, Y.; Entin-Wohlman, O.; David, T.; Salem,
  R.; Folman, R.; Schmiedmayer, J. Long-range order in electronic transport
  through disordered metal films, \emph{Science}  \textbf{2008},
  \emph{319}~(5867), 1226--1229.

\bibitem{Bastard}
Bastard, G. \emph{Wave mechanics applied to semiconductor heterostructures};
  les editions de physique, 1988.

\bibitem{graphene}
Geim, A.K. Graphene: Status and Prospects, \emph{Science}  \textbf{2009},
  \emph{324}, 1530.

\bibitem{judd2011quantum}
Judd, T.E.; Scott, R.G.; Martin, A.M.; Kaczmarek, B.; Fromhold, T.M. Quantum
  reflection of ultracold atoms from thin films, graphene and semiconductor
  heterostructures, \emph{New J. Phys.}  \textbf{2011}, \emph{13}~(8), 083020.

\bibitem{PRB47_2233}
Efros, A.L.; Pikus, F.G.; Burnett, V.G. Density of states of a two-dimensional
  electron gas in a long-range random potential, \emph{Phys. Rev. B}
  \textbf{1993}, \emph{47}~(4), 2233--2243.

\bibitem{pepper}
Berggren, K.F.; Pepper, M. Electrons in one dimension, \emph{Philosophical
  Transactions of the Royal Society of London A: Mathematical, Physical and
  Engineering Sciences}  \textbf{2010}, \emph{368}~(1914), 1141--1162.

\bibitem{OurNJP}
Judd~\textit{et al.}, T.E. {Zone-plate focusing of Bose-Einstein condensates
  for atom optics and erasable high-speed lithography of quantum electronic
  components}, \emph{New J. Phys.}  \textbf{2010}, \emph{12}, 063033.

\bibitem{PhysRevA.83.021401}
Sinuco-Le\'on, G.; Kaczmarek, B.; Kr\"uger, P.; Fromhold, T.M. {Atom chips with
  two-dimensional electron gases: Theory of near-surface trapping and
  ultracold-atom microscopy of quantum electronic systems}, \emph{Phys. Rev. A}
   \textbf{2011}, \emph{83}, 021401.

\bibitem{PhysRevB.92.245439}
Jahn, J.P.; Munsch, M.; B\'eguin, L.; Kuhlmann, A.V.; Renggli, M.; Huo, Y.;
  Ding, F.; Trotta, R.; Reindl, M.; Schmidt, O.G.; et~al. {An artificial Rb
  atom in a semiconductor with lifetime-limited linewidth}, \emph{Phys. Rev. B}
   \textbf{2015}, \emph{92}, 245439.

\bibitem{0953-4075-46-24-245502}
Laycock, T.; Olmos, B.; Montgomery, T.; Li, W.; Fromhold, T.M.; Lesanovsky, I.
  {Control of atomic Rydberg states using guided electrons}, \emph{J. Phys. B:
  Atom., Mol. and Opt. Phys.}  \textbf{2013}, \emph{46}~(24), 245502.

\bibitem{Henkel1999}
Henkel, C.; P{\"o}tting, S.; Wilkens, M. Loss and heating of particles in small
  and noisy traps, \emph{Applied Physics B}  \textbf{1999}, \emph{69}~(5),
  379--387. \urlprefix\url{https://doi.org/10.1007/s003400050823}.

\bibitem{PhysRevA.80.032901}
Contreras~Reyes, A.M.; Eberlein, C. {Casimir-Polder interaction between an atom
  and a dielectric slab}, \emph{Phys. Rev. A}  \textbf{2009}, \emph{80},
  032901.

\bibitem{sernelius2015casimir}
Sernelius, B.E. {Casimir effects in systems containing 2D layers such as
  graphene and 2D electron gases}, \emph{Journal of Physics: Condensed Matter}
  \textbf{2015}, \emph{27}~(21), 214017.

\bibitem{ion_beam2}
Wieck, A.D.; Ploog, K. {In-plane-gated quantum wire transistor fabricated with
  directly written focused ion beams}, \emph{App. Phys. Lett.}  \textbf{1990},
  \emph{56}~(10), 928--930.

\bibitem{ion_beam3}
Ensslin, K.; Petroff, P.M. {Magnetotransport through an antidot lattice in
  GaAs-${\mathrm{Al}}_{\mathit{x}}$${\mathrm{Ga}}_{1\mathrm{-}\mathit{x}}$As
  heterostructures}, \emph{Phys. Rev. B}  \textbf{1990}, \emph{41}~(17), 12307.

\bibitem{xray}
Koonen, J.J.; Buhmann, H.; Molenkamp, L.W. {Probing the Potential Landscape
  Inside a Two-Dimensional Electron Gas}, \emph{Phys. Rev. Lett.}
  \textbf{2000}, \emph{84}~(11), 2473--2476.

\bibitem{EPJD32_171}
Schumm, T.; Est\'eve, J.; Figl, C.; Trebbia, J.B.; Aussibal, C.; Nguyen, H.;
  Mailly, D.; Bouchoule, I.; Westbrook, C.I.; Aspect, A. Atom chips in the real
  world: the effects of wire corrugation, \emph{Eur. Phys. J. D}
  \textbf{2005}, \emph{32}, 171.

\bibitem{RMP79_235}
Fort\'agh, J.; Zimmermann, C. Magnetic microtraps for ultracold atoms,
  \emph{Rev. Mod. Phys.}  \textbf{2007}, \emph{79}, 235--289.
  \urlprefix\url{https://link.aps.org/doi/10.1103/RevModPhys.79.235}.

\bibitem{PRA56_2451}
Sukumar, C.V.; Brink, D.M. Spin-flip transitions in a magnetic trap,
  \emph{Phys. Rev. A}  \textbf{1997}, \emph{56}~(3), 2451--2454.

\bibitem{PhysRevA.72.042901}
Scheel, S.; Rekdal, P.K.; Knight, P.L.; Hinds, E.A. Atomic spin decoherence
  near conducting and superconducting films, \emph{Phys. Rev. A}
  \textbf{2005}, \emph{72}, 042901.
  \urlprefix\url{https://link.aps.org/doi/10.1103/PhysRevA.72.042901}.

\bibitem{PRL97_070401}
Skagerstam, B.S.K.; Hohenester, U.; Eiguren, A.; Redkal, P.K. Spin decoherence
  in superconducting atom-chips, \emph{Phys. Rev. Lett.}  \textbf{2006},
  \emph{97}, 070401.

\bibitem{leanhardt2002propagation}
Leanhardt, A.; Chikkatur, A.; Kielpinski, D.; Shin, Y.; Gustavson, T.;
  Ketterle, W.; Pritchard, D. Propagation of Bose-Einstein condensates in a
  magnetic waveguide, \emph{Physical review letters}  \textbf{2002},
  \emph{89}~(4), 040401.

\bibitem{Mico}
MacLeod, S.J.; Chan, K.; Martin, T.P.; Hamilton, A.R.; See, A.; Micolich, A.P.;
  Aagesen, M.; Lindelof, P.E. Role of background impurities in the
  single-particle relaxation time of a two-dimensional electron gas,
  \emph{Phys. Rev. B}  \textbf{2009}, \emph{80}, 035310.

\bibitem{APB74_469}
Reichel, J. {Microchip traps and Bose-Einstein condensation}, \emph{App. Phys.
  B}  \textbf{2002}, \emph{74}, 469--487.

\bibitem{SinucoPhDThesis}
Sinuco-Le\'on, G. {Quantum properties of Bose-Einstein condensates coupled to
  semiconductor heterojunctions}. Ph.D. Thesis, University of Nottingham,
  September, 2010.

\bibitem{APL41_277}
Drummond, T.J.; Kopp, W.; Morkoc, H.; Keever, M. {Transport in modulation-doped
  structures
  ${\mathrm{Al}}_{\mathit{x}}$${\mathrm{Ga}}_{1\mathrm{-}\mathit{x}}$As/GaAs
  and correlations with Monte Carlo calculations.}, \emph{Appl. Phys. Lett.}
  \textbf{1982}, \emph{41}, 277.

\bibitem{EPJD7_361}
Thywissen, J.H.; Olshanii, M.; Zabow, G.; Drndi{\'c}, M.; Johnson, K.S.;
  Westervelt, R.M.; Prentiss, M. Microfabricated magnetic waveguides for
  neutral atoms, \emph{Eur. Phys. J. D}  \textbf{1999}, \emph{7}~(3), 361--367.

\bibitem{PRA71_063619}
G\"unther, A.; Kemmler, M.; Kraft, S.; Vale, C.J.; Zimmermann, C.; Fort\'agh,
  J. Combined chips for atom optics, \emph{Phys. Rev. A}  \textbf{2005},
  \emph{71}~(6), 063619.

\bibitem{EPL81_56004}
Roux, C.; Emmert, A.; Lupascu, A.; Nirrengarten, T.; Nogues, G.; Brune, M.;
  Raimond, J.; Haroche, S. {Bose-Einstein condensation on a superconducting
  atom chip}, \emph{Eur. Phys. Lett.}  \textbf{2008}, \emph{81}, 56004.

\bibitem{PhysRevLett.98.260407}
Mukai, T.; Hufnagel, C.; Kasper, A.; Meno, T.; Tsukada, A.; Semba, K.; Shimizu,
  F. Persistent Supercurrent Atom Chip, \emph{Phys. Rev. Lett.}  \textbf{2007},
  \emph{98}~(26), 260407.

\bibitem{PhysRevLett.114.113003}
Weiss, P.; Knufinke, M.; Bernon, S.; Bothner, D.; S\'ark\'any, L.; Zimmermann,
  C.; Kleiner, R.; Koelle, D.; Fort\'agh, J.; Hattermann, H. Sensitivity of
  Ultracold Atoms to Quantized Flux in a Superconducting Ring, \emph{Phys. Rev.
  Lett.}  \textbf{2015}, \emph{114}, 113003.

\bibitem{1367-2630-17-11-112002}
Horsley, A.; Du, G.X.; Treutlein, P. Widefield microwave imaging in alkali
  vapor cells with sub-100 μ m resolution, \emph{New Journal of Physics}
  \textbf{2015}, \emph{17}~(11), 112002.
  \urlprefix\url{http://stacks.iop.org/1367-2630/17/i=11/a=112002}.

\bibitem{bohi2010imaging}
B{\"o}hi, P.; Riedel, M.F.; H{\"a}nsch, T.W.; Treutlein, P. Imaging of
  microwave fields using ultracold atoms, \emph{App. Phys. Lett.}
  \textbf{2010}, \emph{97}~(5), 051101.

\bibitem{High_mob}
Umansky, V.; Heiblum, M.; Levinson, Y.; Smet, J.; N\"ubler, J.; Dolev, M. {MBE
  growth of ultra-low disorder 2DEG with mobility exceeding $35 \times 10^6$
  cm$^2$/Vs}, \emph{J. Crystal Growth}  \textbf{2009}, \emph{311}, 1658.

\bibitem{Fleischmann}
Topinka, M.A.; LeRoy, B.J.; Westervelt, R.M.; Shaw, S.E.J.; Fleischmann, R.;
  Heller, E.J.; Maranowski, K.D.; Gossard, A.C. Coherent branched flow in a
  two-dimensional electron gas, \emph{Nature}  \textbf{2001}, \emph{410}, 183.

\bibitem{Buks2}
Buks, E.; Heiblum, M.; Shtrikman, H. Correlated charged donors and strong
  mobility enhancement in a two-dimensional electron gas, \emph{Phys. Rev. B}
  \textbf{1994}, \emph{49}~(20), 14790.

\bibitem{Coleridge}
Coleridge, P.T. Correlation lengths for scattering potentials in
  two-dimensional electron gases, \emph{Semicond. Sci. Technol.}
  \textbf{1997}, \emph{12}, 22.

\bibitem{Grill}
Grill, R.; D\"ohler, G.H. {Effect of charged donor correlation and Wigner
  liquid formation on the transport properties of a two-dimensional electron
  gas in modulation $\delta$-doped heterojunctions}, \emph{Phys. Rev. B}
  \textbf{1999}, \emph{59}, 10769.

\bibitem{doi:10.1080/00031305.1988.10475524}
Rodgers, J.L.; Nicewander, W.A. Thirteen Ways to Look at the Correlation
  Coefficient, \emph{The American Statistician}  \textbf{1988}, \emph{42}~(1),
  59--66.

\bibitem{siegert}
Siegert, C.; Ghosh, A.; Pepper, M.; Farrer, I.; Ritchie, D.A. The possibility
  of an intrinsic spin lattice in high-mobility semiconductor heterostructures,
  \emph{Nature Physics}  \textbf{2007}, \emph{3}~(5), 315--318.

\bibitem{1107.2268}
Haakh, H.R.; Henkel, C. Magnetic near fields as a probe of charge transport in
  spatially dispersive conductors, \emph{The European Physical Journal B}
  \textbf{2012}, \emph{85}~(1), 46.
  \urlprefix\url{http://dx.doi.org/10.1140/epjb/e2011-20567-1}.

\bibitem{SST9_2031}
Buks, E.; Heiblum, M.; Levinson, Y.; Shtrikman, H. Scattering of a
  two-dimensional electron gas by a correlated system of ionized donors,
  \emph{Semicon. Sci. Technol.}  \textbf{1994}, \emph{9}, 2031.

\bibitem{PRB52_11284}
Palm, T. Effects of remote impurity scattering including donor correlations in
  a branching electron waveguide, \emph{Phys. Rev. B}  \textbf{1995},
  \emph{52}, 11284.

\bibitem{hinds}
Hinds, E.A.; Hughes, I.G. Magnetic atom optics: mirrors, guides, traps, and
  chips for atoms, \emph{Journal of Physics D: Applied Physics}  \textbf{1999},
  \emph{32}~(18), R119.

\end{thebibliography}

\appendix

\section{Current flow in high-mobility 2DEGs}
\label{sec:2DEG}

\begin{SCfigure}[50.0][!htb]
  \centering
  \includegraphics[width=9.5cm]{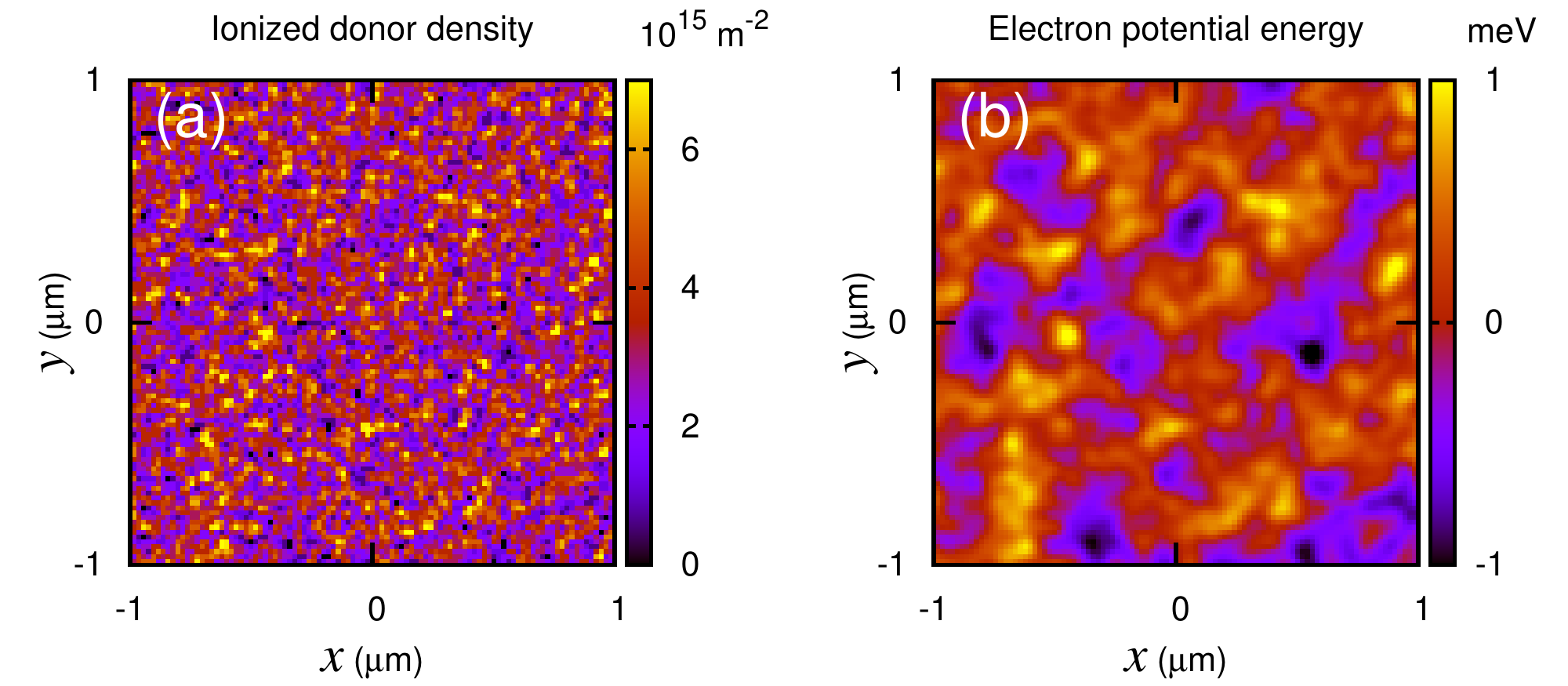}
  \caption{\label{two_deg_distribution}(a) Typical random distribution of ionised donor density calculated in the plane of the $\delta-$doping layer, with donor density $n=3.3\times10^{15}~$m$^{2}$, (b) Corresponding screened  electrostatic potential energy of an electron in the 2DEG layer, located at a distance $d=52.9$nm. Scales shown right.} 
\end{SCfigure}

We consider a  GaAs/(AlGa)As heterojunction with layer structure as shown in figure \ref{geometry}. A  2DEG (blue layer) is formed by electrons from donors in a Si $\delta$-doping layer (red), which migrate into the GaAs and populate the ground state of an almost triangular potential well formed at the GaAs/(AlGa)As interface.  This confines the electrons in a narrow ($\sim$ 15 nm thick) sheet and leaves them free to move in a plane parallel to the GaAs/(AlGa)As interface \cite{PRB47_2233}. The  heterojunction  contains a  layer of ionised donors of mean density  $n = 3.3 \times 10^{15}~$m$^{-2}$, which  is  located at a distance  $d = 52.9$ nm from a 2DEG with the same density (figure \ref{two_deg_distribution}) \cite{PRB47_2233}.  The 2DEG is $20$ nm below the bottom surface of the heterojunction and its mobility is $\mu = 140~$m$^2$V$^{-1}$s$^{-1}$.   To operate in the high electron mobility regime, the heterojunction must be kept at a temperature around $\sim 4.2~$K, similar to that of a superconducting atom chip \cite{PhysRevLett.98.260407,EPL81_56004}. 

The motion of electrons in the 2DEG is affected by the background layer of ionised donors, whose distribution depends on factors like the fabrication process, illumination and thermal history \cite{xray,pepper}.  In the high-mobility regime, the electron mean free path, i.e. the average distance travelled by an electron before being scattered by an impurity or defect, can be  much  larger than the characteristic length scale of the inhomogeneities in the potential landscape, through which the electron moves, which originate from non-uniformity of the ionised donors \cite{SST9_2031,Buks2,Grill}. In $\delta$-doped heterojunctions, the distance between the donors and the 2DEG planes determines the length scale of the potential fluctuations, since features with smaller characteristic lengths are exponentially suppressed (see equation (\ref{screened_potential}) below) \cite{JAP65_361}.   

Provided that the 2DEG is within an electric field, $E_0$, which is large enough to create current streamlines that are only weakly perturbed by the potential fluctuations, the current flow and the potential landscape are related by Ohm's law:
\begin{equation}
\boldsymbol{j}(x,y) = \sigma E_0 \boldsymbol{\hat{x}} + \sigma
\boldsymbol{\nabla}_{2D}  \Phi(\boldsymbol{x,y}),
\label{eq:current_density}
\end{equation}
where $\sigma = n e \mu$ is the mean 2DEG conductivity and $\Phi(x,y)$  is the electrostatic potential experienced by an electron in the 2DEG. In the Thomas-Fermi screening model \cite{PRB47_2233}, $\Phi(x,y)$ is given by:    
  
\begin{equation}
  \Phi(x,y) = \frac{e^2}{4 \pi \epsilon \epsilon_0} \int  e^{-kd}
  \frac{\Delta n (\boldsymbol{k}) e^{i(k_x x + k_y y)}}{k + k_s} d^2\boldsymbol{k}
  \label{screened_potential}
\end{equation}
where $\textbf{k} = (k_x,k_y)$, $k=\left|\textbf{k}\right|$, $\Delta n (\textbf{k})$ is the 2D Fourier transform of the spatial ionised donor density fluctuations from the mean, $\Delta n (x,y)$, $\epsilon=12.9$ is the relative permittivity of GaAs, and the screening wave vector, $k_s = e^2m^*/(2 \epsilon \epsilon_0 \pi \hbar^2)$, depends on the electron effective mass, $m^*$ \cite{PRB52_11284,PRB47_2233}.  Figures  \ref{two_deg_distribution}(a) and (b), respectively, show a typical ionised donor distribution and the corresponding electrostatic potential energy, $\Phi(x,y)$, of a 2DEG with mean density $n=3.3\times10{^{15}}~$m$^{2}$ and $k_s=2.1\times10^{8}~$m$^{-1}$.

\section{Magnetic field produced by currents in a 2DEG}
\label{sec:field}
The model presented in Appendix \ref{sec:2DEG} focuses on the effects of inhomogeneity of the ionised donor distribution on the current flow in high-mobility 2DEGs. In turn, the characteristics of the magnetic field produced by such a flow,  such as its rms amplitude and characteristic length scale, can also be related to the donor distribution.  Figure \ref{two_deg_properties_fig_and_B_x_at}(a) shows a typical  potential energy landscape calculated for an electron in the 2DEG, including current stream lines (black). Figures \ref{two_deg_properties_fig_and_B_x_at}(b)-(d), respectively, show the corresponding variation of $B_x(\boldsymbol{r},z)$ in the $z=1~\mu$m, $3~\mu$m and $5~\mu$m planes parallel to the 2DEG.   Note that these magnetic field fluctuations are independent of the electric field applied to the 2DEG, provided that this electric field is large enough to create current streamlines that are only weakly perturbed by the potential energy fluctuations

At distances from the 2DEG larger than the correlation length of the donor distribution ($\sim 10$ nm), the length scale of variations in the magnetic field landscape is dominated by the distance, $z$, from the 2DEG \cite{APL88_264103}. Understanding the variation of the magnetic field landscape with this distance is crucial for designing devices that couple the 2DEG to nearby atoms. Here, we evaluate the spatial average of the component of the magnetic field, $B_x$, parallel to the mean electron flow, which can be measured via BEC magnetometry \cite{APL88_264103,Nature_wildermuth} and is given by:
\begin{eqnarray}
\centering
(B_{x}^{\text{rms}}(z))^2 =   \left( \frac{\mu_0
  \sigma e}{4 \epsilon \epsilon_0} \right)^2  \int \int
d^2\boldsymbol{k}d^2\boldsymbol{k'} 
  \frac{k_{y} k'_y \left\langle
    \Delta n(\boldsymbol{k})\Delta n (\boldsymbol{k'})\right\rangle}{(k +
    k_s)(k'+k_s)}  &\times& \nonumber  e^{-(k+k')(d+z)} \\ &\times&  e^{i(k_x + k'_x)x}.
\label{B_x_rms}
\end{eqnarray}

Equation (\ref{B_x_rms}) is a key result, it shows us that the $B_x^{\text{rms}}$ dependence with $z$ is shaped by the correlation function of the ionised donor density. This enables us to reduce the amplitude of  $B_x^{\text{rms}}$ by tailoring the donor statistics, which can be achieved by thermal cycling, etching, ion deposition  or illuminating the heterojunction \cite{xray}. In this last case,  the ionised donor distribution can be permanently altered and patterned by transiently illuminating the device with a laser standing wave. Such static and dynamical control of local carrier density does not exist for metallic current-carrying conductors whose geometric and material-related field inhomogeneity can only be reduced by continously applied time-dependent external fields \cite{PRL98_263201}. 

In general, semiconductor fabrication procedures (e.g. molecular beam epitaxy) produce homogeneous and isotropic donor distributions, whose  correlation function  depends only on the relative distance between two points, $|\boldsymbol{r}-\boldsymbol{r'}|$. Correspondingly, in Fourier space, $S(\boldsymbol{k},\boldsymbol{k'})$ is proportional to $\delta(\boldsymbol{k}+\boldsymbol{k}_x)$, which means that equation (\ref{B_x_rms}) reduces to: 
\begin{equation}
(B_x^{\text{rms}}(z_0))^2 = \left( \frac{\mu_0
  \sigma e}{4 \epsilon \epsilon_0} \right)^2 \frac{ <\Delta n ^2>}{(2 \pi)^2}  
  \pi k_s^2 \int d\tilde{k}    \tilde{k}^3  
  \frac{1}{(\tilde{k} + 1)^2} e^{-2\tilde{k}(d+z)k_s}, 
\label{int_b_rms}
\end{equation}
where $\tilde{k} = k/k_x$ and  $\left\langle \Delta n^2\right\rangle$ is the mean-square average of the ionised donor density spatial fluctuations. In this form, the integrand in equation (\ref{int_b_rms}) is dimensionless, and its numerical evaluation yields a power law decay as function of ($d + z_0$), specifically:
\begin{equation}
( B_x^{\text{rms}}(z))^2 = \left( \frac{\mu_0  \sigma e}{4 \epsilon
     \epsilon_0} \right)^2 \frac{<\Delta n^2>}{(2 \pi)^2} \pi
   \frac{2.17 \times 10^{-10}}{(d+z)^4}  ,
\label{B_x_rms_final}
\end{equation}
with $(d+z_0)$ in microns.

We now consider the decay of the magnetic field fluctuations produced by a heterostructure whose ionised donor density varies periodically, with period $\lambda = 2\pi/k_0$, along the $y-$direction. In this case,
\begin{equation}
S(\boldsymbol{k},\boldsymbol{k}') = \frac{\delta n^2}{4\pi^2}\delta(k_y-k_0) \delta(\boldsymbol{k}+\boldsymbol{k}'),
\end{equation}
where we have assumed an ionised donor density modulation of amplitude $\delta n$.

After integrating Eq. (\ref{B_x_rms}), the dependence of the rms amplitude of the magnetic field with the distance to the chip becomes:
\begin{equation}
(B_x^{\text{rms}}(z))^2 = \left( \frac{\mu_0
  \sigma e}{4 \epsilon \epsilon_0} \right)^2 (\delta n)^2 \frac{4k_0^2}{(2k_0 + k_s^2)^2}e^{-4k_0(d+z_0)}.
\label{eq:expDecay}
\end{equation}

This demonstrates that periodic modulation of the donor distribution creates an exponential decay of the rms amplitude of the magnetic field fluctuations, similar to the magnetic mirror in  \cite{hinds}. Such a  modulation can be created permanently by etching or implanting Ga ions \cite{ion_beam3} or,  by optical transient illumination of the sample with a periodic laser standing wave pattern \cite{xray}. 

Figure \ref{B_x_decay_BS} in the main text compares $B_x^{\text{rms}}(z)$ calculated using Eq. (\ref{eq:expDecay}) (solid/red curve) and Eq. (\ref{B_x_rms_final}) (dashed/blue curve) along with the corresponding field fluctuations for a metal wire including surface and edge fluctuations \cite{EPJD32_171} (short-dashed/black curve). Insets in the same figure show the effect of the ionised donor density distribution on the potential energy landscape of electrons in the 2DEG with (left-hand inset) and without (right-hand inset) periodic modulation. Crucially, exponential decay makes the $B_{x}^{\text{rms}}(z_0)$ curve for the periodically-modulated donor distribution rapidly fall off to a value below that for a metal wire. By patterning the donor distribution with a period of $~200$ nm, at $z \gtrsim 0.5~\upmu$m, the field fluctuations above the 2DEG are more than 3 orders of magnitude  smaller than for the metal wire. 
                                 
The ability to tailor the potential landscape of the 2DEG, and the resulting field fluctuations, is a unique feature of heterojunctions and can be exploited for trapping, manipulating, and imaging with, ultracold Bose gases. 

\end{document}